# Effects of Network Connectivity and Functional Diversity Distribution on Human Collective Ideation

**Authors:** Yiding Cao[1†*], Yingjun Dong[1‡], Minjun Kim[1,2], Neil G. MacLaren[1,3,4], Sriniwas Pandey[1], Shelley D. Dionne[1,3], Francis J. Yammarino[1,3], and Hiroki Sayama[1,3,5*]

**Affiliations:**

[1]Binghamton Center of Complex Systems, Binghamton University, State University of New York, Binghamton, NY, USA.

[2]State University of New York at Plattsburgh, Plattsburgh, NY, USA.

[3]Bernard M. and Ruth R. Bass Center for Leadership Studies, Binghamton University, State University of New York, Binghamton, NY, USA.

[4]Department of Mathematics, State University of New York at Buffalo, Buffalo, NY, USA.

[5]Waseda Innovation Lab, Waseda University, Tokyo, Japan.

*Correspondence to: ycao20@binghamton.edu and sayama@binghamton.edu.

[†]Current address: Michigan Medicine, University of Michigan, Ann Arbor, MI, USA.

[‡]Current address: University of Texas Health Science Center, Houston, TX, USA.

**Abstract:**

Human collective tasks in teams and organizations increasingly require participation of members with diverse backgrounds working in networked social environments. However, little is known about how network structure and the functional diversity of member backgrounds would interact with each other and affect collective processes. Here we conducted three sets of human-subject experiments which involved 617 university students who collaborated anonymously in a collective ideation task on a custom-made online social network platform. We found that spatially clustered collectives with assortative background distribution tended to explore more diverse ideas than in other conditions, whereas collectives with random background distribution consistently generated ideas with the highest utility. We also found that higher network connectivity may improve individuals' overall experience but may not improve the collective performance regarding idea generation, idea diversity, and final idea quality.

## Introduction

Organizations are increasingly relying on diverse collectives for successful solution development for many real-world technical and business problems *(1-5)*. In such collectives, multiple people with different backgrounds work together to achieve a greater goal than would be possible for individuals to accomplish alone *(6-9)*. Sharing of expertise among collective members with diverse backgrounds and behaviors is recognized as an important factor in collective effectiveness *(10-14)*. Researchers are hence interested in how to improve the quality and efficiency of human collective processes in organizational settings *(15,16)*. Human collective processes involve interaction and interdependence among multiple individuals with task-related functional diversity in expertise, experience, knowledge, and skills (note that these are different from demographic or identity diversity as often discussed in public media and literature) *(17-20)*. Some research has indicated that the multidisciplinary backgrounds within collectives are positively correlated with quantitative and qualitative task performance *(7,10-12,19-23)* and that social network structure influences collective performance *(19-29)*, although results obtained so far have been rather mixed and fragmented. This is because there are several difficulties in investigating human collective processes *(27)*, including complex organizational structure *(30,31)*, open-endedness of problems/tasks *(31)*, heterogeneity of participants *(32)*, and long-term dynamic nature of the processes *(33,34)*. It has thus been difficult to conduct experimental investigation of the effects of (and the interactions among) these complex factors altogether on human collective processes in realistic collaborative settings. Accordingly, previous studies were often computer simulation-based *(7,14,15,17-20,26,29)* or, if experimental, mostly limited in terms of collective size, organizational structure, duration of collaboration, and/or complexity of tasks *(18,27,35,36)*.

This study aims to address the above gap in the existing literature. Specifically, we experimentally investigated how the functional diversity of backgrounds of individual members and their social network structure interact with each other and affect human collective ideation processes in realistic settings with a larger collective size, a longer collaboration period, and a more open-ended ideation task for which no simple solutions would exist. The key methodological improvements we newly developed in this study include (1) a custom-made online collaboration platform on a configurable social network that allowed for detailed experimental control and data collection, and (2) the use of modern machine learning tools for natural language processing and quantitative semantic analysis that were used in both experimental manipulation and resulting data analysis.

Three sets of experiments were conducted using a custom-made online social network platform. In each experimental session, 20~25 anonymous participants were arranged to form a social network according to their backgrounds and collaborated on text-based collective ideation tasks for two weeks. The performance of collective ideation was characterized using multiple metrics, including the number of generated ideas, the best/average quality score of final submitted ideas, semantic diversity of generated ideas quantified using machine learning-based word embedding methods, and post-experiment survey results on participants' overall experience. By analyzing these performance metrics of collectives with various network configurations, we aim to address the following research questions: (1) How does the participants' background distribution within a collective affect the performance of collective ideation and the participants' experience? (2) How does the network structure/density of a collective affect the performance of collective ideation and the participants' experience?

We designed and built a custom-made online social network platform with a Twitter/X-like user interface for operating the experiments (see "Experimental platform" section in Methods). This platform allows participants to submit ideas in response to the assigned collaboration task, and to discuss the task by reading, commenting on, and liking other participants' ideas. The experiments involved a total of 617 participants who were undergraduate or graduate students majoring in different disciplines at a mid-sized US public university. To participate in an experimental session, students were required to fill out an experimental registration form to provide information about their academic major and a (relatively long, about 300-word) written description of why they selected their major, as well as their academic knowledge, technical skills, career interest areas, hobbies, or extracurricular activities, and/or any other information related to their background (this narrative information is called "background" in this study). The background information was used to characterize participants' functional diversity and to determine their arrangement within the social network in each experimental session.

Each online experimental session lasted for 10 working days (= 2 weeks), during which participants were assigned an anonymous username to log in to the experimental online platform and spend about 15 minutes each day to work on the collective ideation task by submitting ideas and commenting on and liking their collaborators' ideas. Participants were expected to continuously elaborate on and improve their ideas over time by utilizing their collaborators' ideas and comments provided on the platform. After the experimental session was over, the participants were asked to submit an end-of-the-session survey form to provide their favorite final ideas. These final ideas were later evaluated on a 5-point Likert scale by multiple third-party experts who were not involved in the experiments (Ph.D. candidates majoring in Marketing or Management for the first "laptop slogan" task and professionals working in the University's Communications division for the second "story writing" task; they both possessed strong writing and evaluation skills). These evaluation scores provided by all evaluators were finally averaged and then used to quantitatively assess the utility values of the final ideas (i.e., idea quality) made by each collective. The end-of-the-session survey also included four questions about the participants' overall experience in the experiment, level of learning and understanding from their collaborators, self-evaluation of own contribution to the collaborative process, and personal evaluation of the final ideas (see Methods for details). This survey was conducted in order to evaluate the participants' experience during the experimental session and examine how it might be affected by the network configuration.

In this study, we conducted three sets of controlled online human-subject experiments. Each experiment was named according to the network structure of participants' collaborative relationships and the type of the collective ideation task the participants worked on. The first experiment was named the "spatially clustered-slogan design" experiment, abbreviated as "SC-SL" in which 'SC' represents spatially clustered network structure of the collective and 'SL' represents the 'slogan' design task used in this experiment. The second experiment was named the "SC-ST" experiment, which was with spatially clustered network structure and the 'story' writing task. The third experiment was named the "FC-SL" experiment, in which 'FC' represents fully connected (complete) network structure of the collective and 'SL' indicates that it used the slogan design task as in the SC-SL experiment. The slogan design task was considered as a high-collaboration task which involved a lot of interactions among participants, whereas the story writing task was considered as a low-collaboration task which typically involved much less interaction among participants in comparison with the slogan design task. Both tasks involved complex cognitive processes in creative writing, and therefore it was anticipated that functional diversity could play some effects in those collective processes, particularly in diversifying the ideas generated. See Methods for more details of these experimental tasks.

The SC-SL experiment was composed of 4 sessions and the SC-ST experiment was composed of 3 sessions. In each session of the SC-SL and SC-ST experiments, participants were split into three separate collectives, which were configured to be similar to each other in terms of the network size, network structure, and the amount of within-collective background variations (i.e., average distance of backgrounds between participants). These numerical vectors representing the background characteristics of the participants were used, together with the participants' academic majors, to arrange the participants within the social network. The underlying social network structure was a spatially clustered regular network made of 20−25 members with a node degree of four (Fig. 1A-C). This network structure is equivalent to a Watts-Strogatz small-world network with no random wiring *(37)*. Participants connected to each other were able to observe each other's posts and activities on the online platform but would not directly see activities of other nonadjacent participants.

The three collectives in each SC session differed only regarding spatial distributions of participants' background variations, which is called "background distribution" hereafter. We tested three different background distribution conditions: assortative (Condition A, Fig. 1A) in which participants were connected to other participants with similar backgrounds; random (Condition R, Fig. 1B) in which participants were connected randomly regardless of their backgrounds; and disassortative (Condition D, Fig. 1C) in which participants were connected to other participants with distant backgrounds.

Finally, the FC-SL experiment was composed of 2 sessions with a total of 8 fully connected collectives with similar network size of collectives to those in SC-SL and SC-ST sessions (Fig. 1D). The detailed information of all experimental sessions is shown in Table 1.

The data acquired from the registration forms, experimental idea posts, and the end-of-the-session survey forms were mostly in plain text format, which would be hard to analyze using traditional quantitative analysis methods. Therefore, we converted those text data to numerical vectors using word embedding (also called semantic embedding, sentence embedding or text vectorization) techniques developed and used in Machine Learning and Natural Language Processing. Approaches of word embedding have been applied for measuring expertise diversity and textual idea analysis in other studies *(38, 39)*. We used the Doc2Vec algorithm *(40)* in this study. Doc2Vec, an adaptation of Word2Vec *(41)*, is an unsupervised machine learning algorithm that can generate numerical vectors as a representation of sentences, paragraphs, or documents. Compared to other algorithms, Doc2Vec can provide a better text representation with a lower prediction error rate, because it can recognize the word ordering and semantics of words which are not accounted for by other algorithms *(42)*.

The written descriptions of backgrounds submitted by participants in the registration forms for each experimental session were used to train background Doc2Vec models for that session. The outputs of these background Doc2Vec models were given in the form of 400-dimensional numerical vectors, which were combined with the self-reported academic major information to quantitatively represent the background features of the participants. The numerical background vectors were used as inputs for participant allocation within the spatially clustered network which was described earlier. The daily posted and final design ideas were converted to 100-dimensional numerical vectors using Doc2Vec. The combined set of all the ideas of the SC-SL and FC-SL experimental sessions was used as the corpus for training the "slogan design Doc2Vec" model. The whole collection of ideas of the SC-ST experimental sessions was used as the corpus for training the "story writing Doc2Vec" model. The outputs of these Doc2Vec models were used to quantify the diversity of ideas generated in collective ideation (i.e., average Euclidean distance among the idea vectors).

**Results**

The results of the SC-SL and SC-ST experiments were used to study how the participants' background distribution would affect the collective performance and the participants' experience. Meanwhile, the results of the SC-SL and FC-SL experiments were compared to study how the network structure/density would affect the collective performance and the participants' experience. The results of the FC-SL experiment were also used for preliminary assessment of the relationship between the functional diversity of participants and the diversity of ideas generated by them, which confirmed our anticipation that the functional diversity of participants had positive effects on idea diversification in their collective ideation processes (see Fig. S1 for details).

For the SC-SL experiment, we compared the outcome measures among the three background distribution conditions. The results show that the collectives in Condition A (assortative) made fewer daily posts than the collectives in Condition R (random) (Fig. 2A), but the distances among those posts were significantly greater than those generated by the collectives in Condition R or D (disassortative) (Fig. 2C). Moreover, in all the four SC-SL sessions, the collectives with Condition R always generated the best final idea with the highest utility value (Fig. 3A). Meanwhile, in all the four SC-SL sessions, the collectives with Condition D consistently achieved the highest average utility score of the final ideas (Fig. 3B). As there were only four sessions run in this experiment, we would not be able to derive a statistically significant conclusion from this result. However, the consistent patterns observed across the three background distribution conditions imply the possibility of such organizational arrangements to have impacts on collective ideation and innovation.

The results of SC-ST experiment show that the collectives in Condition R made fewer daily posts than the collectives in Condition A or D (Fig. 2B) but the distances among those posts were not statistically different among the three conditions (Fig. 2D). The latter finding was particularly interesting as it shows a stark contrast with the results of the SC-SL experiment (Fig. 2C). The effect of background distributions on the idea diversity is clearly manifested in the high-collaboration tasks (e.g., the slogan design task in this study) but may not be so if the collective process does not involve much collaborative interactions among participants.

Finally, the results of FC-SL show that there was no significant difference between spatially clustered and fully connected collectives regarding the ideation activities (numbers of daily posts) (Fig. 4A), but the average distances between generated ideas were significantly less in fully connected networks (Fig. 4B). Moreover, there appeared to be a moderate level of difference in the highest score (Fig. 4C) and the average score (Fig. 4D) of final ideas, indicating that high connectivity in fully connected networks may not help to improve the quality of ideas. These findings suggest that high density of a social network is not necessarily a positive factor to improve idea generation, idea exploration or idea quality in collective collaboration.

We also analyzed the participants' answers regarding the end-of-the-session survey questions to evaluate their experience during the experiment. The results of SC-SL and SC-ST show that the difference in background distribution conditions did not significantly affect the participants' survey responses, but significant differences were found between spatially clustered collectives in the SC-SL sessions and fully connected collectives in the FC-SL sessions regarding the self-evaluated overall quality of ideas and the learning experience (Fig. 5). This is probably because of the greater number of ideas each participant was exposed to in a fully connected social environment in the FC-SL experiment. Note that these improved subjective experiences of the participants were not aligned with the objective performance metrics of the collectives shown in Fig. 4. It would require more in-depth investigation (possibly with another set of experiments) to derive possible mechanistic explanations for why the participants in the fully connected networks gained improved subjective experience, which is beyond the scope of this study.

**Discussion**

In this study, we conducted a series of controlled online human-subject experiments to investigate the effect of background distribution and network structure on collective ideation processes. The results revealed non-trivial interactions among the social network structure, the distribution of functional diversity and the nature of collaborative tasks in realistic settings, which were not reported before. Specifically, the diversity of generated ideas was significantly greater when the network structure was of low density and spatially clustered and when the background distribution on the network was assortative. Interestingly, this observation was obtained only for the experimental sessions with the slogan design task but not for the sessions with the story writing task, indicating that the effects of background assortativity were on the collaborative interactions among collective members. This result can be understood in that the assortativity of participants' backgrounds helps different parts of the collective explore possible solutions in different directions and thus diversify the results of collective exploration *(20)*. This is analogous to biological diversity promoted and maintained in spatially structured evolutionary populations *(19)*. It was also notable that our participant survey results indicated that the participants gained significantly better experience and more satisfaction when the network was dense and fully connected, even though the actual performance of the collective ideation was actually lower. This presents an important lesson that the perceived and actual performances of collective ideation may not be correlated with each other.

This study also indicated the potential difference of collective performance between different background distribution conditions in spatially clustered regular networks. When participants were randomly placed, the collective tended to find the best ideas most effectively. This seemingly puzzling observation may be understood by considering how much background diversity each generated idea was exposed to locally. Namely, in Condition A, generated ideas are exposed to human participants that were relatively homogeneous, and thus those ideas only need to meet relatively simple, consistent criteria to be successful in spreading. In Condition D, in contrast, ideas are exposed to and evaluated by very different human participants, and thus those ideas must satisfy a wide variety of (possibly inconsistent) criteria, necessarily making them conservative and mistake-proof. These two situations may be analogized to the widely discussed "exploration" (variation-driven dynamics in creative conditions) vs. "exploitation" (selection-driven dynamics in critical conditions) spectrum *(18-20,26)*. We hypothesize that collectives in Condition R achieved the right balance in the middle of this exploration vs. exploitation spectrum and thereby found the best ideas most frequently, and meanwhile, that collectives in Condition D had a high ability to filter out potentially problematic ideas and generate ideas that can be commonly accepted by most participants, achieving the highest average score. These results and interpretations altogether paint a picture of the complex nature of human collectives; the structure and composition of a collective with regard to functional diversity of participating individuals should be considered and designed according to the objectives and success measures.

This study is still limited in various aspects. First and foremost, the number of experimental sessions was rather small, and hence we were unable to obtain statistically significant results for several collective-level outcome measurements (e.g., highest utility and average utility achieved in each session). This limitation was largely due to the complexity of the experiments. Replicating the same experiments for a larger number of times would require a substantial effort and cost. Second, in this study, the measurement of functional diversity and idea diversity was almost entirely based on the Doc2Vec machine learning algorithm. There are many other possible methods to characterize and quantify those diversities, ranging from more conventional human coding to more recent Large Language Model (LLM)-based methods. Furthermore, the design tasks we used in the experiments were simple text-based creative tasks, and therefore it remains unclear whether and how our findings would generalize to other kinds of tasks, such as visual/auditory creative tasks, more analytical tasks such as solving mathematical/logical puzzles, and more critical tasks such as evaluating a business plan. Testing the robustness/sensitivity of the key findings under these method/task variations will be an important and informative next step of this study.

## Methods

### *Information of experimental sessions*

We conducted three sets of controlled online human-subject experiments: SC-SL (Spatially Clustered-SLogan design), SC-ST (Spatially Clustered-STory writing), and FC-SL (Fully Connected-SLogan design). The experiment name is shown in Table 1 with session numbers and conducting time aligned with each session. The network size (number of participants) and background distribution of collectives in each session were also recorded in this table.

### *Collective ideation tasks*

The ideation tasks used in the experiments were unknown to participants before the experiment began. We created and used two open-ended textual design tasks with no obvious solutions immediately available to any participants. The task used in the SC-SL and FC-SL experiment was a task which asked participants to create slogans, taglines, or catch phrases for marketing a laptop. This task typically involved a lot of interactions among participants, in which new ideas were often built on discussion, advocation, and innovation of existing ideas proposed by the participants. In contrast, the task used in the SC-ST experiment was a task which asked participants to write a short story or a complete fiction within a character count limit. This task, compared to the first one, typically involved much less interaction among participants. The full description of these two tasks is shown in Table 2.

### *Experimental platform*

The human-subject experiments were conducted fully online using our original, custom-made web-based computer-mediated collaboration (CMC) platform. This platform was developed using Flask, a Python framework for simple, efficient, extensive web application development *(43)*. The interface of this platform (shown in Fig. 6) is minimalistic, but its functionality is similar to that of some social media websites such as Twitter/X. This experimental platform provides a web-based toolkit for experiment administrators to implement various experimental settings and run/manage online experiments. Basic information of participants, including their real names, anonymized account names, login credentials, and email addresses, can be entered to the system by uploading a csv file of the participants' registration information to the system's administrative page. The platform also allows the administrators to post and update a task description and start/end experimental sessions through the backend of the system. It is also easy to configure the structure of collectives (including network structure and background distribution) by uploading a network adjacency list file to this platform, in which participants are connected to each other based on a pre-configured network structure whose global shape is not visible to the participants. Most importantly, the platform provides an intuitive, precise, and efficient way for the experimental administrators to monitor participants' activities and download the data of participants' posts (generated ideas).

To work on the collaboration task during the experimental session, participants can log in to the experimental server using their username and password. For new users, they can check basic instructions from the Help page shown in Fig. 6(A). When participants work on their task, they can always find the collective ideation task description displayed at the top of the platform website. Through this platform, they can browse the current set of candidate ideas in the timeline, create a new idea (either by coming up with an entirely novel idea or by utilizing and modifying existing ones) and submit it to the platform to share it with others. They can also comment on existing ideas to show advocacies or criticisms of these posts or click "Like" on them to show support as well (Fig. 6(B)). The contact information of the experimental administrators is shown at the bottom of the page to help participants ask questions whenever they have any and report any system errors.

***Experimental procedure and protocol***

To participate in an experimental session, participants were required to fill out an informed consent and experimental registration form (created and administered using Google Forms) to provide their current academic major and a written description of their background in about 300 words, including why they selected their major, their previous majors, academic knowledge, technical skills, career interest areas, hobbies or extracurricular activities, and/or any other information related to their background. These statements were analyzed using the Doc2Vec word embedding algorithm *(40)* to quantify semantic contents and their distances in a "background space" among the participants (see "Word embedding" below for more details). These inputs from participants were used to measure and control their background diversity for participant grouping. When registering for the experiment, participants' email addresses and some other private information were recorded. To protect the privacy of participants, such information was used only for communications related to this project during the experiment and they were removed after the whole experiment procedure was over. In the registration form, participants could also find information about the overall objective of this research, the experimental procedure, and the experimental protocols. The whole human subject experimental protocol was reviewed and approved by the Binghamton University Institutional Review Board (IRB).

Before an experimental session began, participants received a notification email which provided the instruction of how to participate in the experiment with the platform server's URL (IP address), their randomly generated login username and password, as well as the starting and ending dates of the experimental session. Each online experimental session lasted for two weeks (10 working days), during which participants were requested to log in to the experimental platform and spend at least 15 minutes on each working day on the assigned collective ideation task by collaborating with their anonymous neighbors in the social network constructed in the experimental system. Their participation and collaboration actions were logged electronically in the experimental server and monitored by the experiment administrators on a regular basis.

On each working day during the experimental session, a daily reminder email was sent to each participant with the reminders of the experimental server's URL and their login username and password so that they can start their work. They were requested to post one or more novel or modified ideas on the platform each day and discuss the task by reading and commenting on their connected collaborators' ideas. The experimental regulation will only allow participants to submit posts that are in the scope of the collaboration task. Any obvious intrusive posts that are unrelated to the task will be deleted by the experimental administrator during the data collection and cleaning procedure. In fact, only very few posts were considered as irrelevant and/or intrusive ideas which were deleted after the experiment. By potentially utilizing their collaborators' ideas and comments received from their collaborators, participants were expected to continuously elaborate and improve their idea quality over time.

After the two-week experimental session was over, participants received a final reminder email which asked them to complete an end-of-the-session survey form within one week. This end-of-the-session survey asked participants to choose and submit three final ideas from the posting records in their experimental platform timeline. These final ideas were later evaluated by multiple third-party experts who were not involved in the experiments. Those third-party experts were Ph.D. candidates majoring in Marketing or Management for the first "slogan design" task, and professionals working in the University's Communications division for the second "story writing" task. Their rating scores generally showed good agreement. These evaluation scores provided by all evaluators were finally averaged and then used to quantitatively assess the utility values of the final ideas (i.e., idea quality) made by each collective. This survey also included questions about participants' overall experience in the experiment, level of learning and understanding from their collaborators, self-evaluation of own contributions to the collaborative process, and personal evaluation of the final ideas (Table 3).

***Participant recruitment***

We recruited a multidisciplinary group of students through several courses offered at a mid-sized US public university from Fall 2018 to Spring 2020 (academic semesters) to participate in the experiments. Recruited participants were undergraduate or graduate students majoring in Management (Accounting, Business Analytics, Finance, Quantitative Finance, Management Information Systems, Marketing, Supply Chain Management, Leadership and Consulting, Organizational Behavior and Leadership), Engineering (Industrial and Systems Engineering, Systems Science, Electrical Engineering, Mechanical Engineering) and other disciplines (Psychology, Economics, Actuarial Science, Philosophy, Mathematical Science). The experiments were run from Fall 2018 through Spring 2020, over which a series of recruitment campaigns were used to attract students to participate in the experiment. In total, 617 students participated in the study; 423 worked on the slogan design task, and 194 on the story writing task (see Table 1).

Recruitments were conducted through class announcements and email advertisements using IRB-approved recruitment scripts. It was made clear to students in the announcement that participation in this experiment would be entirely voluntary and would not affect their future relations with their class instructor, the university, or the investigators of this research.

Recruitment required participants to be over the age of 18, be able to use a computer keyboard and a pointing device for computer-mediated collaborative work and have such a computer environment to participate. Students who had participated in this series of experiments before were not allowed to register for a second time, in order to minimize the effects of learning. Interested students were guided to the online registration form, where more details of the study, the objective of experiments, and the informed consent were provided. If a student voluntarily agreed to participate in the experiment, they would fill in the online registration form, including their informed consent. They were free to withdraw their consent and discontinue participation at any time without any penalty. Course instructors would not be informed of an individual student's participation until after final grades were assigned.

As an incentive, students who participated in the experiment and completed the entire duration were offered a small amount of extra credit in their course from which they were recruited. Monetary compensation was not offered. After the course instructors assigned final grades at the end of the semester, they were provided the names of students who earned extra credit. The amount of credit a student would earn was uniform for all participants and would not vary based on their performance in the experiment.

During the recruitment, participants were also provided with potential benefits or risks from participating in this experiment. The potential benefits were that they might gain useful collaborative skills and substantive knowledge about collaborative work. Potential risks in participating in this experiment might be the potential negative influence on student's academic work and potential health problems arising from using a computer for a long period of time, both of which would be typical in any student participation in human-subject research. To minimize the risks and ensure the experiment quality, we avoided midterm or final exam weeks, weekends, and holidays in scheduling the experimental sessions. Moreover, the comfortable interface of our CMC experiment platform and the short working time requirements on each day were also considered to minimize the risks.

### *Participants allocation in social network*

In each session of the SC-SL and SC-ST experiments, participants were split into three separate collectives (see Table 1), which were configured to be similar to each other in terms of the network size, network structure, and the amount of within-collective background variations (i.e., average distance of background between participants). The underlying social network was a spatially clustered regular network made of 20~25 members with degree four, in all the collectives in SC-SL and SC-ST sessions (network layout examples are shown in Fig. 1). A spatially clustered regular network is equivalent to a Watts-Strogatz small-world network with no random wiring *(37)*. It should be made clear that these collectives were undirected networks, i.e., participants connected to each other could observe each other's posts and activities in both directions. Meanwhile, participants could not directly see activities of other nonadjacent participants.

The three collectives in each session differed only regarding spatial distributions of participants' background variations. We tested three different background distribution conditions: assortative background distribution (Condition A), random background distribution (Condition R), and disassortative background distribution (Condition D). These background distribution conditions varied the spatial pattern of individual participants' backgrounds within social network structure. The example layouts of three background distribution conditions are shown in Fig. 1A-C in the main text, where participants were represented by nodes colored according to their backgrounds. Similar/different background participants hold similar/different colors, respectively.

The participant allocation described above was done using a computational algorithm implemented in Python. This algorithm would first generate 100 independent trials of splitting participants into three nearly equally sized collectives and structuring them into three spatially clustered networks. For each trial, two measurements were calculated for the generated three collectives: Average Intra-Collective Distance and Average Across-Edge Distance regarding the participants' backgrounds. The Average Intra-Collective Distance (AICD) is the average background difference between every pair of nodes within the network, which represents the overall within-collective background variations. The Average Across-Edge Distance (AAED) is the average background difference between every pair of *connected* nodes within the network, which characterizes the background similarity/dissimilarity between neighbors in the network. These two measurements were acquired by calculating Euclidean distances between background numeric vectors of participants (see "Word embedding" below). Using these measurements, the most optimal trial was selected so that the three collectives had similar AICD values but very different AAED values, i.e., one collective should have a low AAED for Condition A, another collective should have a medium AAED for Condition R, and the last collective should have a high AAED for Condition D.

The FC-SL experiment was composed of 8 collectives with fully connected (complete) network structure in which every pair of participants was connected by an edge (a network layout example is shown in Fig. 1D). The 8 collectives had similar AICD (within-collective background variation), and all were made of 20 participants. The degree of each node in these networks was 19, which was significantly higher than the node degree of spatially clustered networks (just 4) in the SC-SL and SC-ST experiments. Unlike in spatially clustered networks in which individual participants could only observe 4 neighbors' activities, each participant in fully connected collectives could observe all the 19 other participants' activities and might receive more comments and "Like"s.

***Data collection and analysis***

The participants' activity records stored in the experimental server were utilized as the primary dataset for data analysis. The data we obtained from each experimental session include: textual ideas posted by the participants, final ideas selected by the participants, numerical vectors generated from those ideas using the word embedding algorithm (see below), the utility scores of the final ideas evaluated by the third-party experts, and the participants' responses to the end-of-the-session survey questions.

The Mann Whitney U test with the Bonferroni correction was used for most of statistical testing to compare different experimental conditions. The Bonferroni correction was applied to reduce the chances of obtaining false-positive results among multiple conditions *(44)*. In the meantime, for comparison of three background distribution conditions, we used the Wilcoxon signed-rank test instead, which is a paired sample version of the Mann-Whittney U test. The reason of using a paired sample statistical test for this is that each experimental session had recruited rather different participant populations with different backgrounds (e.g., sometimes the majority came from Management; at other times the majority came from Engineering), and thus the equivalence of within-collective background variations was adjusted to match between collectives only within each session, but not across different sessions. The Wilcoxon signed-rank test for paired samples is appropriate in this case as it can rank the absolute values of the differences between the paired observations in Condition A, Condition R, and Condition D within each session and calculates a statistic on the number of negative and positive differences.

***Word embedding***

We used the Doc2Vec implementation of the Gensim Python library. The word embedding procedure starts with text cleaning, including dropping stop words (a set of commonly used short words in English, e.g., “the”, “is”, “and”), removing special characters, removing punctuations, and expanding abbreviations. After that is the tokenization process which is an essential text preprocessing step to separate text into smaller units, called tokens. Doc2Vec will train a model using a set of tokens as corpus, and then use the model to generate vector representation of each text as a numerical output. The length of the output vectors is determined by the size parameter that is set when training the model. Doc2Vec uses a simple neural network with a single hidden layer.

The whole set of tokens obtained from the written descriptions of backgrounds submitted by participants in the registration forms for each experimental session was used as corpus to train background Doc2Vec models of the participants’ backgrounds for that session. The outputs of these Doc2Vec models were given in the form of 400-dimensional numerical vectors, which were combined with the self-reported academic major information to quantitatively represent the background features of the participants.

To investigate whether the Doc2Vec algorithm can effectively capture the background similarity/dissimilarity, we performed a validation test with 30 randomly selected triplets of the written background descriptions extracted from the experimental registration records of one experimental session. For each triplet, one participant (denoted as Participant B) was taken as a reference to separately estimate its background distance with the other two participants (denoted as Participants A and C) to obtain $d_{\mathrm{AB}}$ that represents the background distance between Participants A and B and $d_{\mathrm{BC}}$ that represents the background distance between Participants B and C. For each of the 30 triplets, $d_{\mathrm{AB}}$ and $d_{\mathrm{BC}}$ were estimated by two independent human evaluators as well as by the Doc2Vec algorithm, and their results were compared.

The two human evaluators independently reviewed every pair of the original background descriptions submitted by the experimental participants and rated their background distance using a 5-point Likert scale, ranging from 1 (very similar) to 5 (very distant). For each triplet of background descriptions, if both human evaluators gave the same trend in relative distances among A, B and C (i.e., both gave $d_{AB} < d_{BC}$; both gave $d_{AB} > d_{BC}$; or one gave $d_{AB} = d_{BC}$ and the other gave $|d_{AB} - d_{BC}| \leq 1$), we considered they reached an agreement. The result of this human-based evaluation showed a 100% agreement rate between the two human evaluators, indicating that it would indeed be possible to estimate the participants' background distances (as perceived by natural humans) robustly from the background descriptions collected in this study.

We then compared the human-rated distances (obtained by averaging the distances rated by the two human evaluators) with the Euclidean distances between the 400-dimensional numerical vectors generated from the background descriptions using Doc2Vec. They were found to be positively correlated with strong statistical significance ($p = 0.0084 < 0.01$). Fig. 7 presents a scatter plot showing a clear positive correlation between human-rated distances and Doc2Vec-generated distances. These results provide assurance that one may use Doc2Vec as a reasonable alternative method in place of labor-intensive manual background distance evaluation.

The daily ideas and final ideas were converted to 100-dimensional numerical vectors using Doc2Vec, because the length of ideas was generally much shorter than the length of participants' background descriptions. The combined set of all the ideas of the SC-SL and FC-SL experiments was used as corpus for training the "slogan design" Doc2Vec model. The whole collection of ideas of the SC-ST experiments was used as corpus for training the "story writing" Doc2Vec model. The outputs of these Doc2Vec models were used to quantify the diversity of ideas generated in collective ideation (i.e., average Euclidean distance among the idea vectors).

**Data Availability:** The anonymized experimental dataset (numerical vectors of participants' backgrounds, numerical vectors of generated ideas, and network configurations) is publicly available from the following data repository site: https://orb.binghamton.edu/systems_fac/13/

**Code Availability:** All the software tools developed and used in this study are available from the corresponding authors upon request.

**Acknowledgements:** Authors thank Ankita Kulkarni and Shun Cao for helpful discussions for this study. This work was supported in part by the US National Science Foundation under Grant 1734147 and the JSPS KAKENHI Grant 19H04220.

**Author Contributions:** Yiding Cao: Conceptualization, Methodology, Software, Validation, Formal analysis, Investigation, Data curation, Writing - original draft, Writing - review & editing, Visualization, Project administration. Yingjun Dong: Methodology, Software, Validation, Investigation, Data curation. Minjun Kim: Software, Resources. Neil G. MacLaren: Investigation, Writing - review & editing. Sriniwas Pandey: Software, Validation, Formal analysis, Data curation, Visualization. Shelley D. Dionne: Conceptualization, Writing - review & editing, Supervision, Funding acquisition. Francis J. Yammarino: Conceptualization, Writing - review & editing, Supervision, Funding acquisition. Hiroki Sayama: Conceptualization, Methodology, Software, Resources, Writing - original draft, Writing - review & editing, Supervision, Project administration, Funding acquisition.

**Competing Interests:** Authors declare no competing interests.

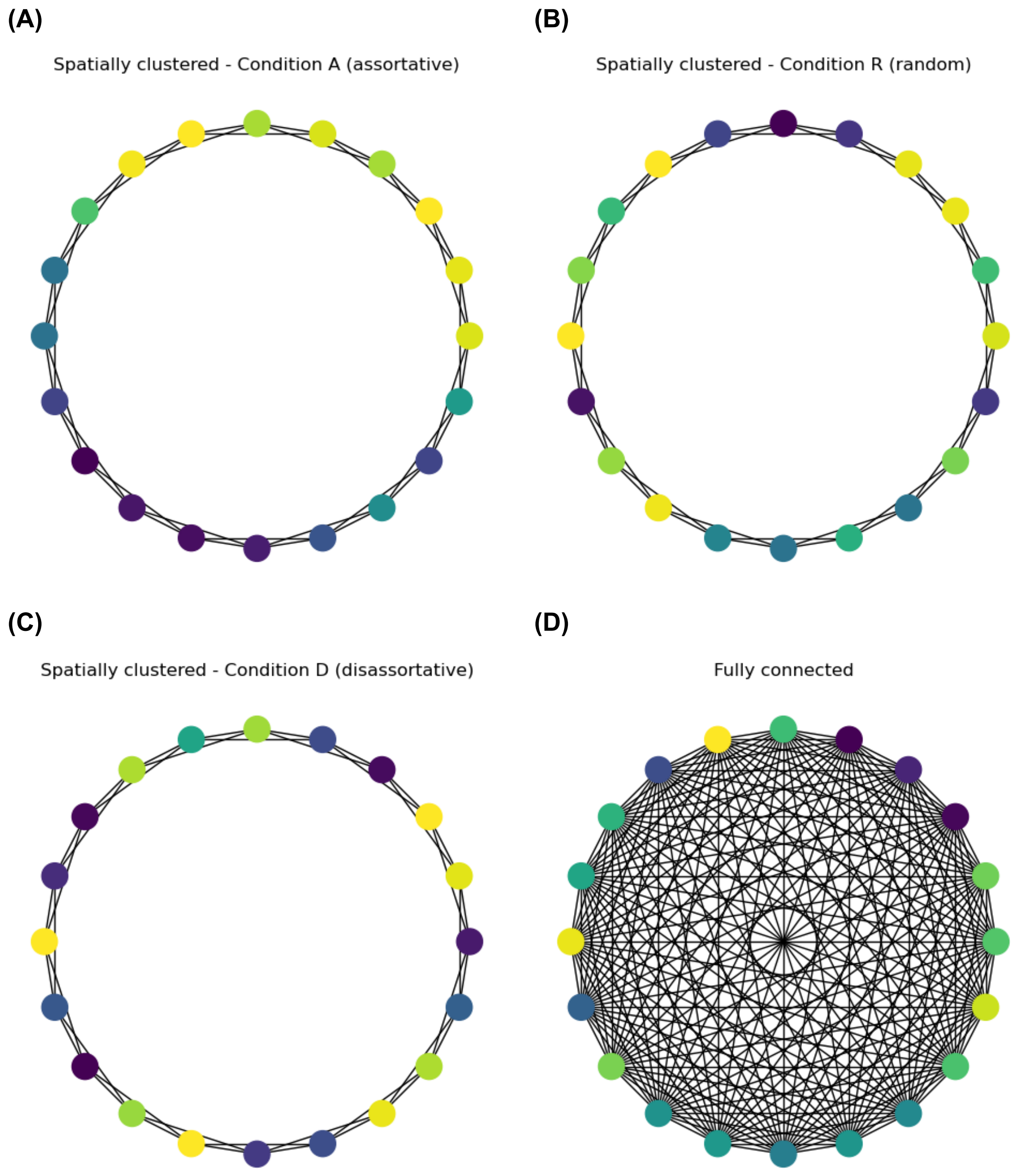

**Fig. 1**
Social network structures used in the experiments. (A) Spatially clustered regular network with background distribution of Condition A (assortative). (B) Spatially clustered regular network with background distribution of Condition R (random). (C) Spatially clustered regular network with background distribution of Condition D (disassortative). (D) Fully connected network. *Note: participants were represented by nodes in the graphs colored according to their backgrounds. Similar/different background participants hold similar/different colors, respectively.*

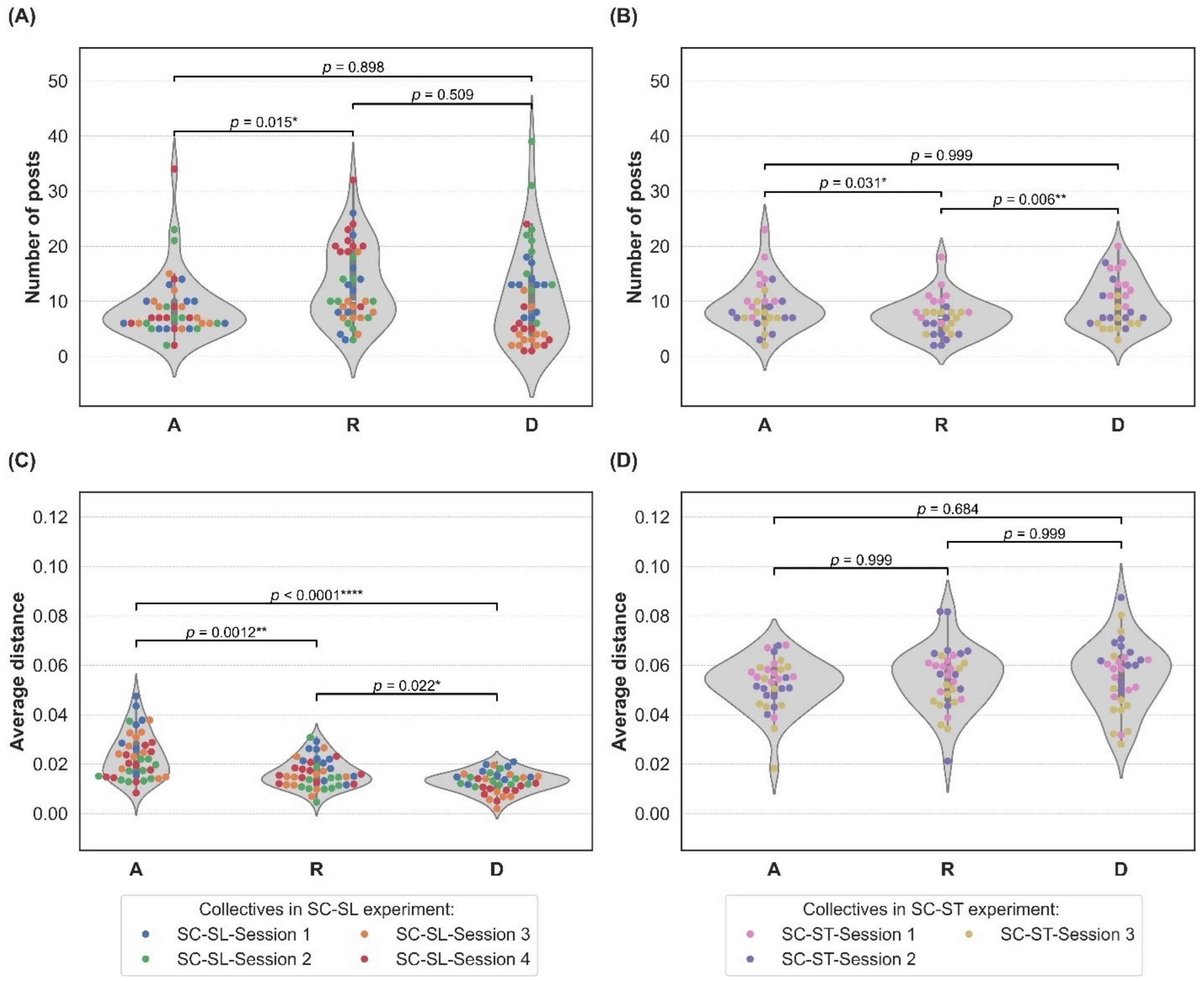

**Fig. 2**

Distribution of numbers of daily posts and average distances between ideas from Day 1 to Day 10. (**A**) Comparison of numbers of daily posts among three background distribution conditions in the SC-SL experiment. (**B**) Comparison of numbers of daily posts among three background distribution conditions in the SC-ST experiment. (**C**) Comparison of average distances between ideas among three background distribution conditions in the SC-SL experiment. (**D**) Comparison of average distances between ideas among three background distribution conditions in the SC-ST experiment. Each dot in these violin plots represents the number of posts (for (**A**) and (**B**)) and the average value of distances between ideas (for (**C**) and (**D**)) of a collective for a single working day. The *p*-value annotation legend is as follows. *: $0.01 < p <= 0.05$, **: $0.001 < p <= 0.01$, ****: $p <= 0.0001$. The Wilcoxon signed-rank test with Bonferroni correction was used for all tests. Data points regarded as outliers were removed in this analysis. Results with all the data points included are shown in Fig. S2.

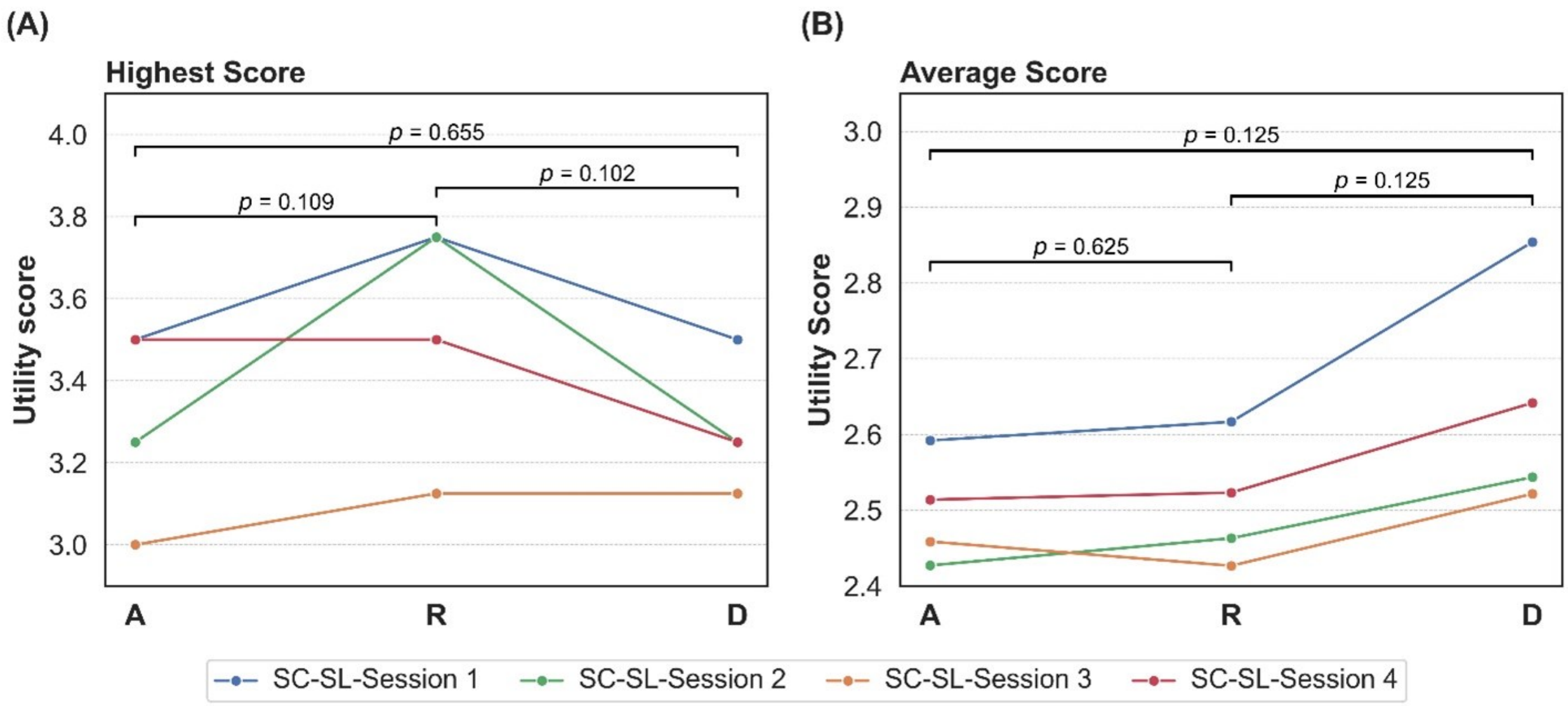

**Fig. 3**
Comparison of idea quality among three background distribution conditions in the SC-SL experiment. (**A**) Highest score of final ideas. (**B**) Average score of final ideas. The Wilcoxon signed-rank test with post-hoc comparison was used for all tests.

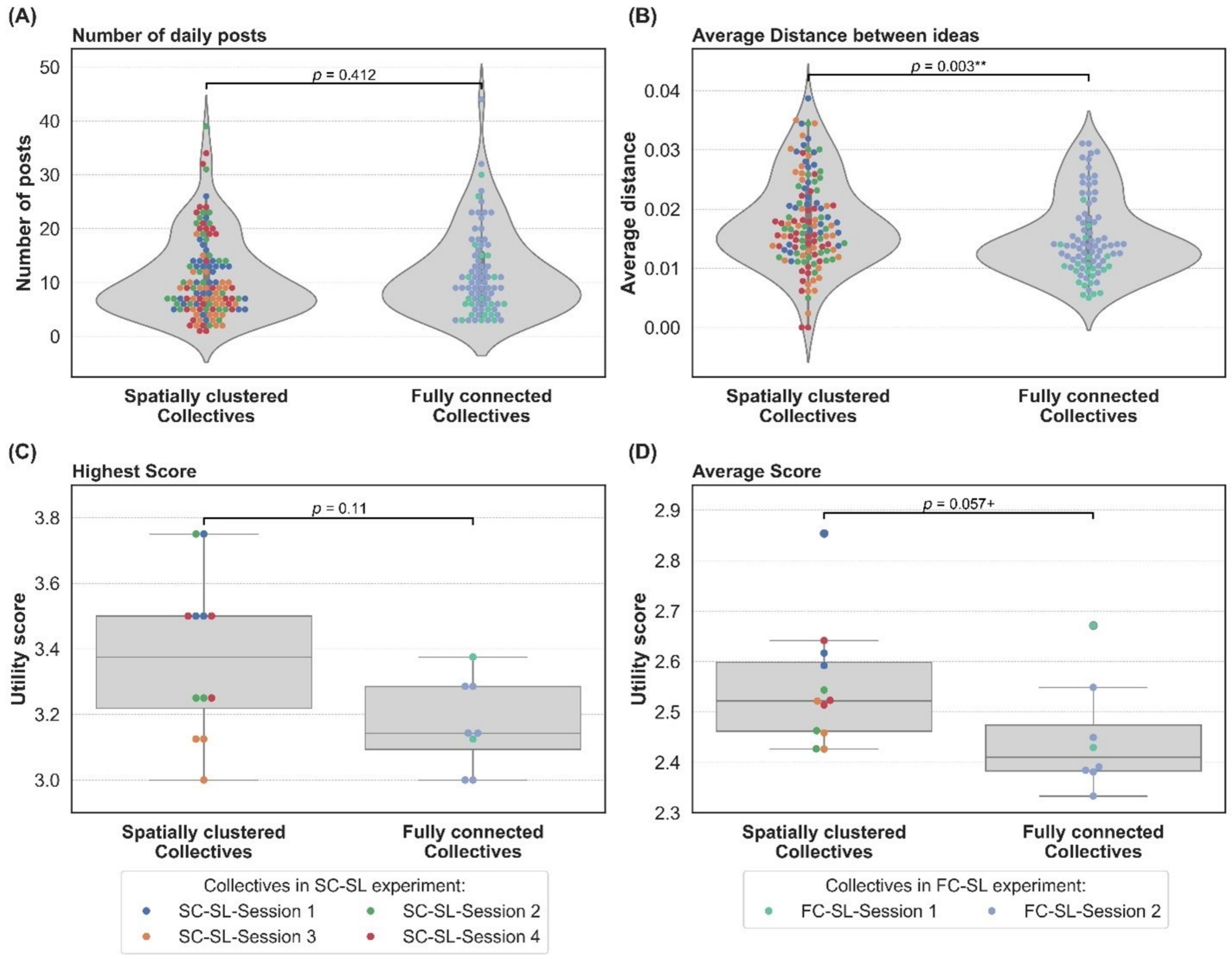

**Fig. 4**
Comparison of results between spatially clustered collectives in the SC-SL experiment and fully connected collectives in the FC-SL experiment. (**A**) Number of daily posts. (**B**) Average distances between ideas. Each dot in these violin plots represents the number of posts (for (**A**)) and the average value of distances between ideas (for (**B**)) of a collective for a single working day. (**C**) Highest score of final ideas. (**D**) Average score of final ideas. The *p*-value annotation legend is as follows. +: $0.05 < p <= 0.1$, **: $0.001 < p <= 0.01$. The two-sided Mann-Whitney U test was used for all tests. Data points regarded as outliers were removed in this analysis. Results with all the data points included are shown in Fig. S3.

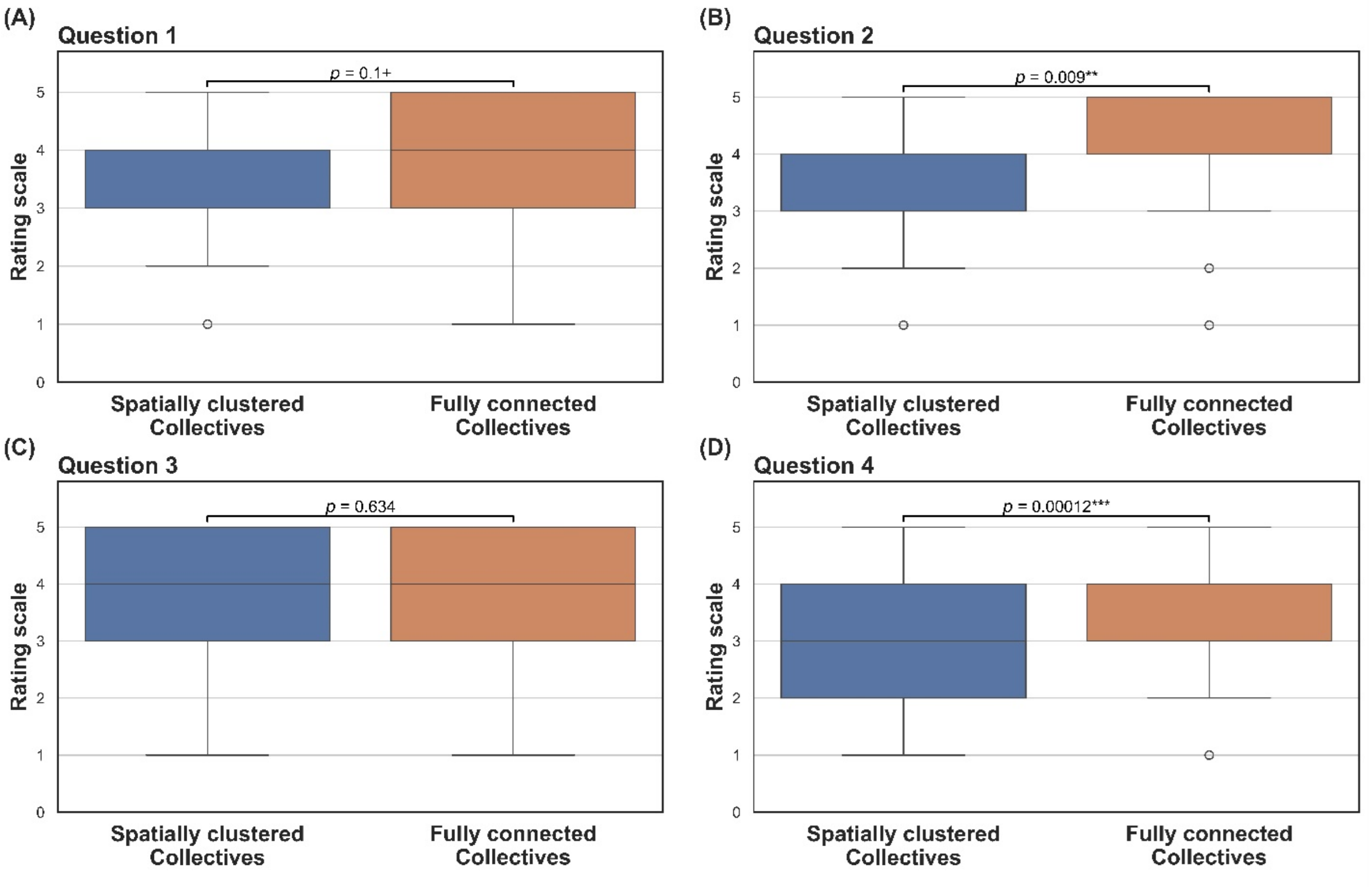

**Fig. 5**
Comparison of rating scales regarding the end-of-session survey questions between spatially clustered and fully connected collectives. (**A**) Question 1 about overall experience. (**B**) Question 2 about self-evaluated overall quality. (**C**) Question 3 about self-evaluated contribution. (**D**) Question 4 about learning experience. The *p*-value annotation legend is as follows. +: $0.05 < p <= 0.1$, **: $0.001 < p <= 0.01$, ***: $0.0001 < p <= 0.001$. The two-sided Mann-Whitney U test was used for all tests.

(A)

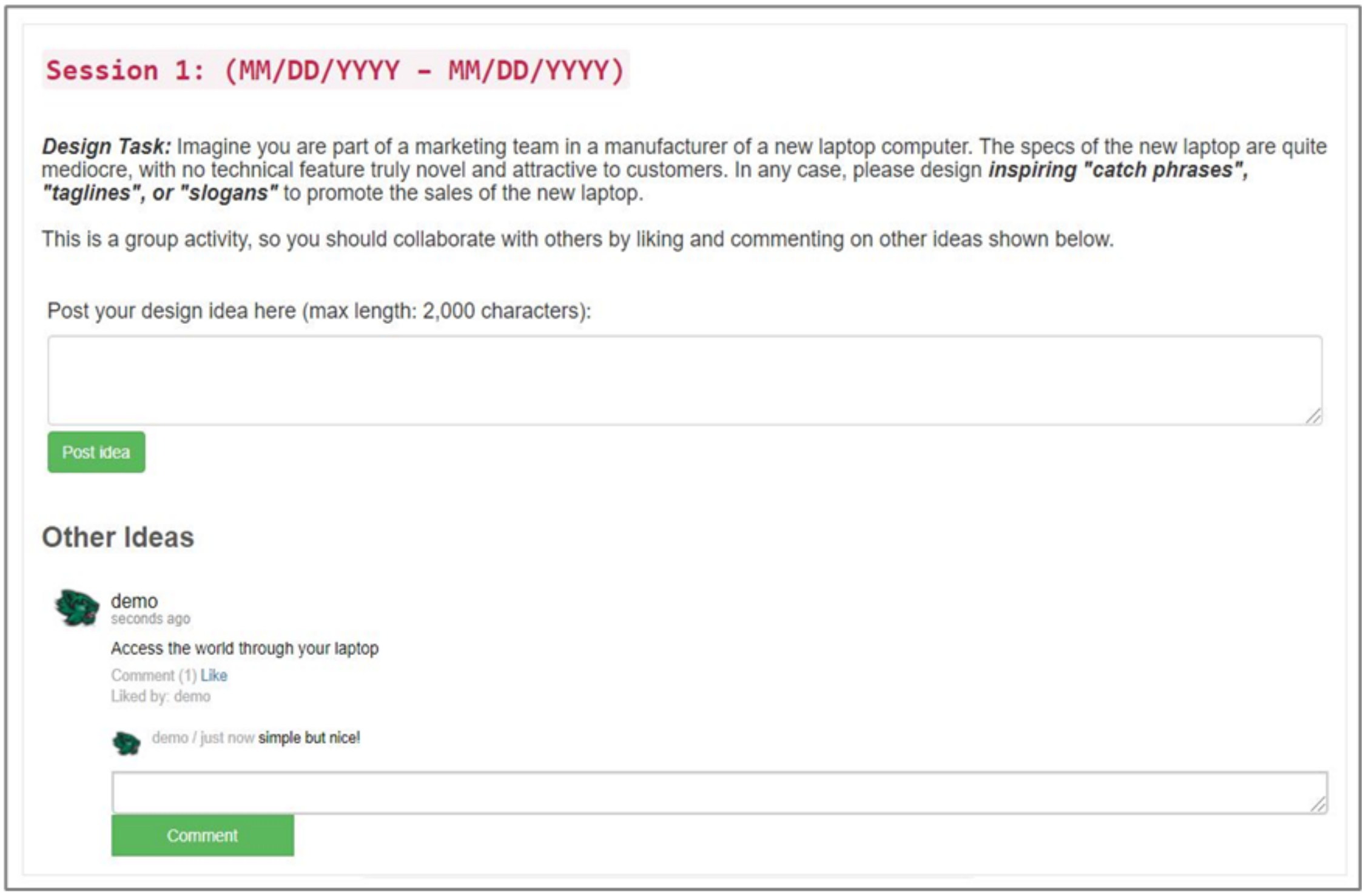

(B)

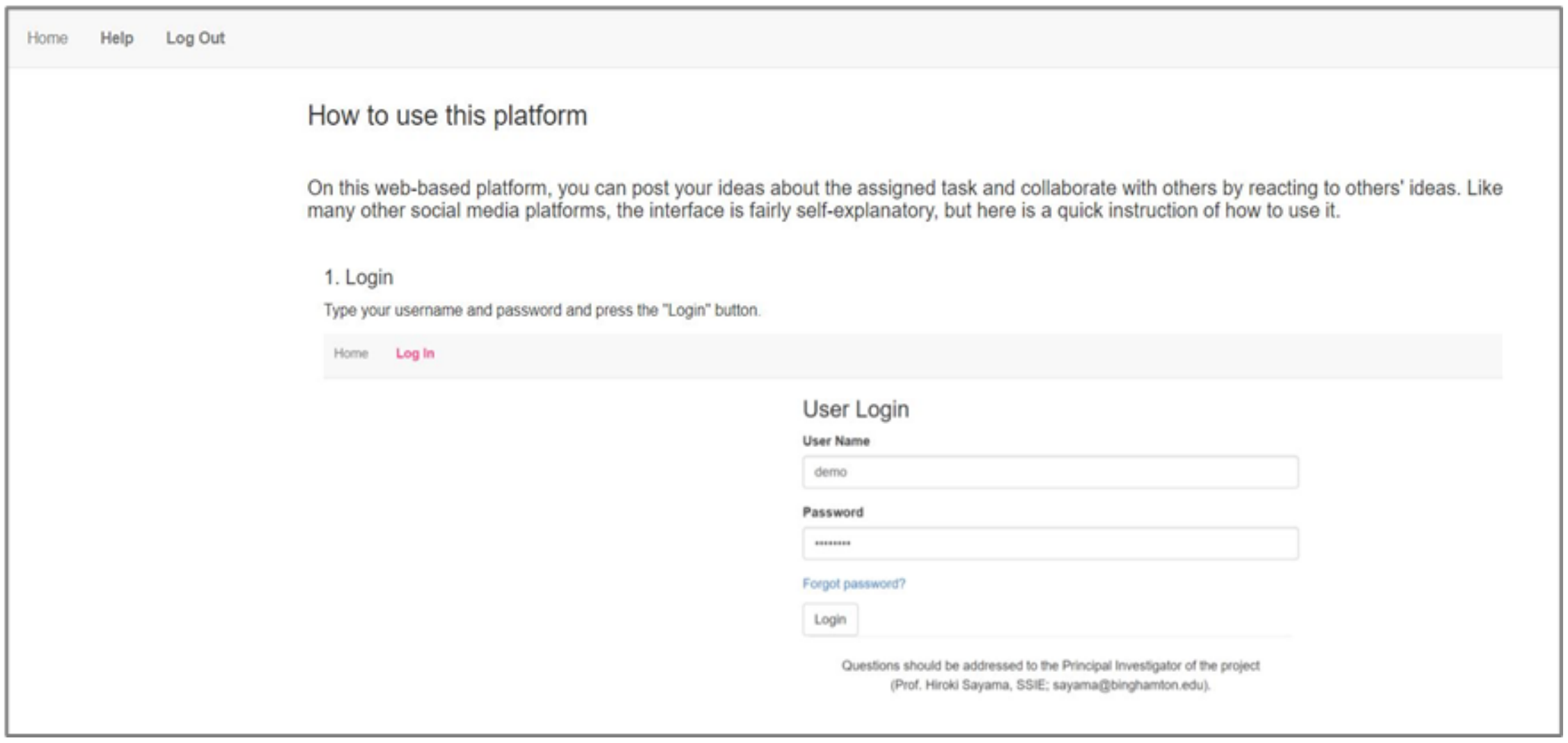

**Fig. 6.** Experimental platform interface. (**A**) The instruction page of the platform. (**B**) The working page of the platform.

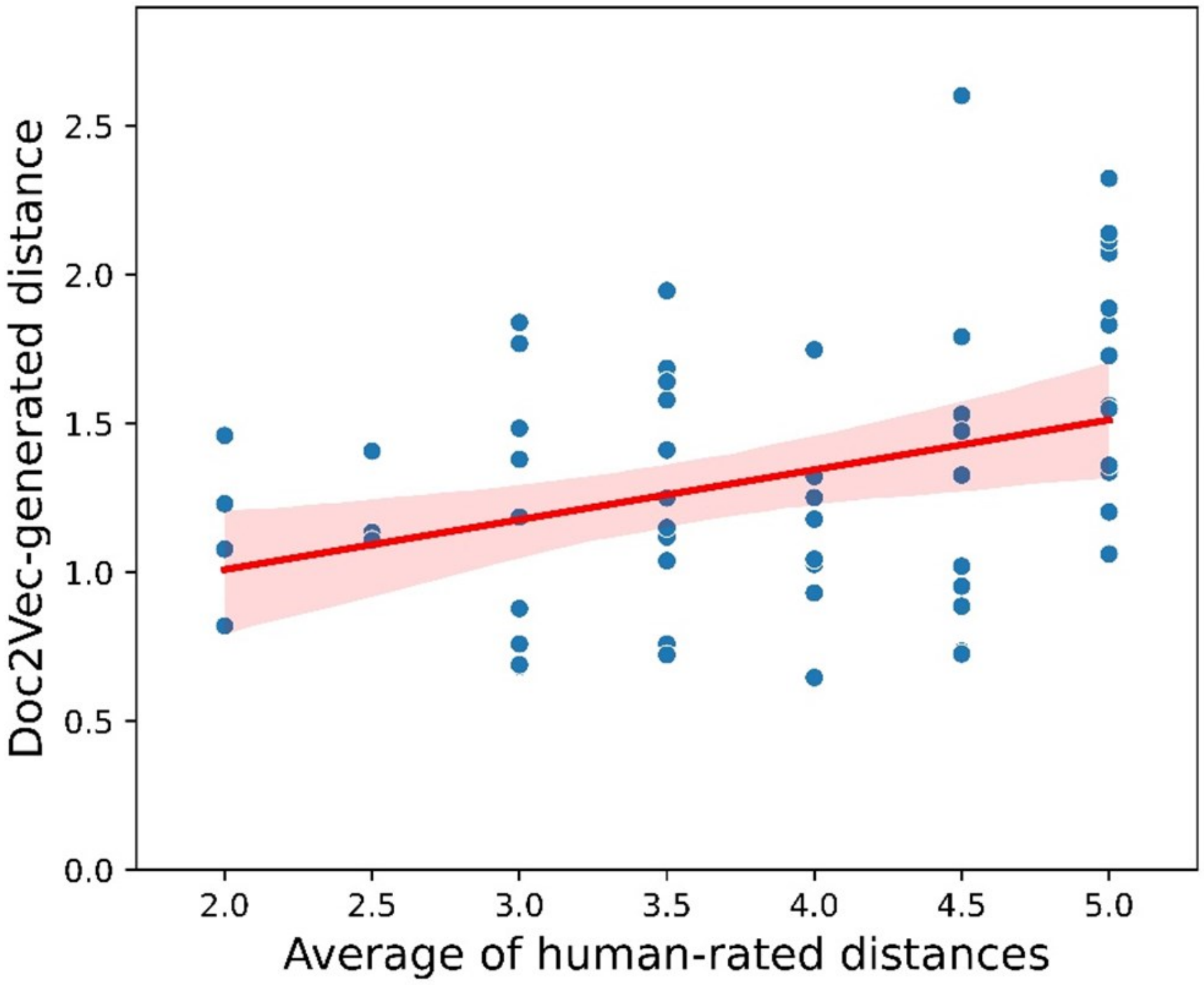

**Fig. 7.** Scatter plot showing significant positive correlation ($p = 0.0084 < 0.01$) between human-rated distances and Doc2Vec-generated distances, together with the linear regression fitting line and its confidence intervals.

| Experiment name | Session number | Dates of experiment | Collective number | Number of participants | Background distribution |
|---|---|---|---|---|---|
| SC-SL experiment | 1 | Fall 2018 | 1 | 22 | A |
| | | | 2 | 20 | R |
| | | | 3 | 22 | D |
| | 2 | Spring 2019 | 1 | 22 | A |
| | | | 2 | 22 | R |
| | | | 3 | 22 | D |
| | 3 | Spring 2020 | 1 | 20 | A |
| | | | 2 | 20 | R |
| | | | 3 | 20 | D |
| | 4 | Spring 2020 | 1 | 25 | A |
| | | | 2 | 24 | R |
| | | | 3 | 24 | D |
| SC-ST experiment | 1 | Spring 2019 | 1 | 25 | A |
| | | | 2 | 24 | R |
| | | | 3 | 25 | D |
| | 2 | Fall 2019 | 1 | 20 | A |
| | | | 2 | 20 | R |
| | | | 3 | 20 | D |
| | 3 | Spring 2020 | 1 | 20 | A |
| | | | 2 | 20 | R |
| | | | 3 | 20 | D |
| FC-SL experiment | 1 | Fall 2019 | 1 | 20 | N/A |
| | | | 2 | 20 | N/A |
| | 2 | Fall 2019 | 1 | 20 | N/A |
| | | | 2 | 20 | N/A |
| | | | 3 | 20 | N/A |
| | | | 4 | 20 | N/A |
| | | | 5 | 20 | N/A |
| | | | 6 | 20 | N/A |

**Table 1.** Information of experimental sessions.

| | |
|---|---|
| **Task 1**<br>**(Slogan design task)** | ***Design Task***: *Imagine you are part of a marketing team in a manufacturer of a new laptop computer. The specs of the new laptop are quite mediocre, with no technical feature truly novel and attractive to customers. In any case, please design* ***inspiring "catch phrases"****,* ***"slogans"****, or* ***"taglines"*** *to promote the sales of the new laptop.*<br><br>*This is a group activity, so you should collaborate with others by liking and commenting on other ideas shown below.* |
| **Task 2**<br>**(Story writing task)** | ***Design Task***: *Write a story. A story can be at any length in general, but for this task, write* ***a complete work of fiction in 280 characters or fewer.***<br><br>*This is a group activity, so you should collaborate with others by liking and commenting on other ideas shown below.* |

**Table 2.** Task descriptions given to participants in the experiments.

| | |
|---|---|
| **Question 1** | How was your overall experience during this session?<br>Rating scale: "Very negative" to "Very positive" (Likert scale: 1 - 5) |
| **Question 2** | How do you feel about the overall quality of the final design(s) you submitted above?<br>Rating scale: "Very poor" to "Very good" (Likert scale: 1 - 5) |
| **Question 3** | How much do you think you have contributed to the collaborative process?<br>Rating scale: "Not at all" to "Quite a lot" (Likert scale: 1 - 5) |
| **Question 4** | How much do you think you have learned about the preferences and behaviors of other participants you communicated with during the session?<br>Rating scale: "Not at all" to "Quite a lot" (Likert scale: 1 - 5) |

**Table 3.** End-of-the-session survey questions.

# Supplementary Materials for

## Effects of Network Connectivity and Functional Diversity Distribution on Human Collective Ideation

**Authors:** Yiding Cao*, Yingjun Dong, Minjun Kim, Neil G. MacLaren, Sriniwas Pandey, Shelley D. Dionne, Francis J. Yammarino, and Hiroki Sayama*

*Correspondence to: ycao20@binghamton.edu and sayama@binghamton.edu.

The following pages present additional plots and visualizations of the experimental results that were not included in the main text.

(A)

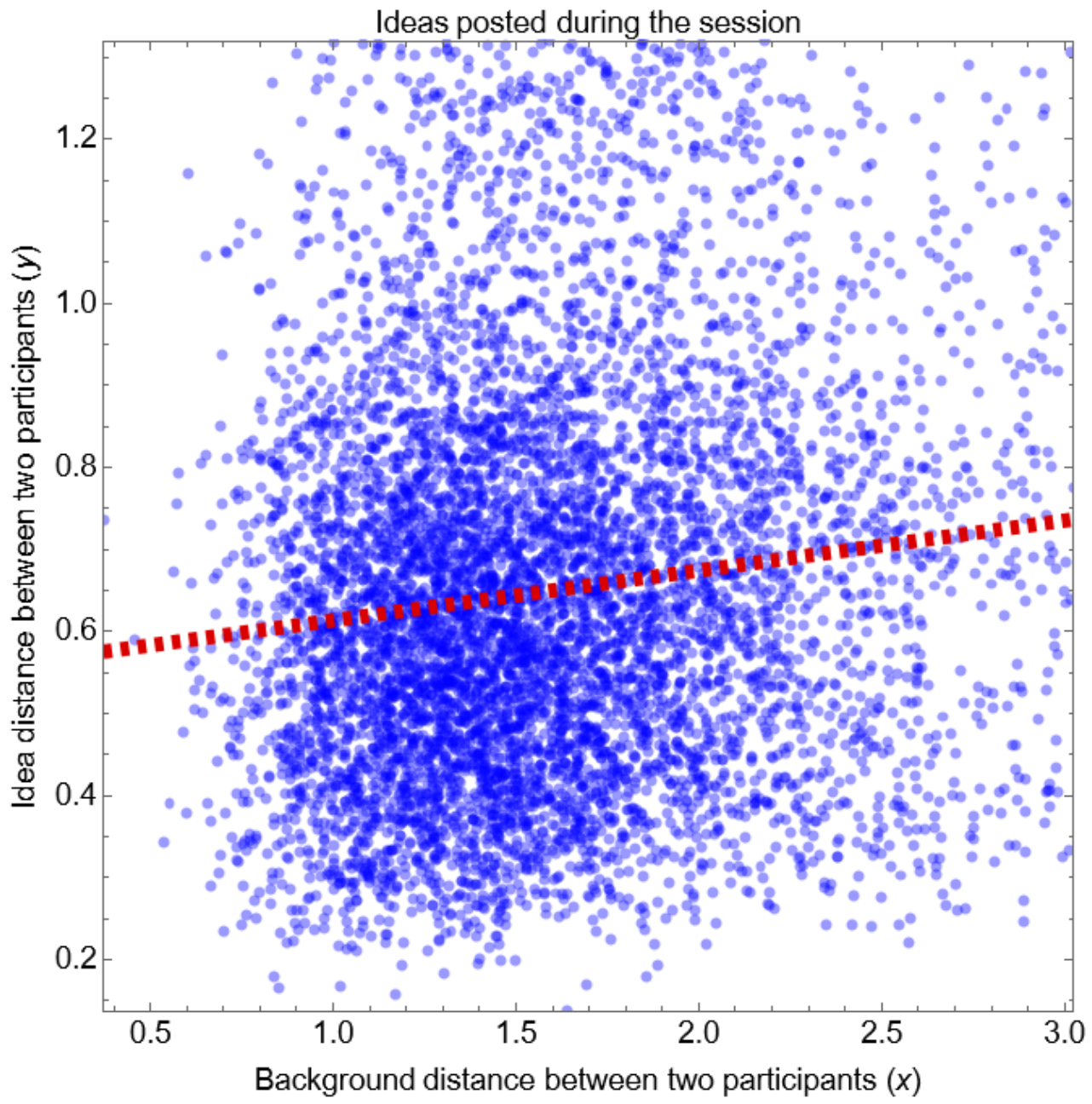

(B)

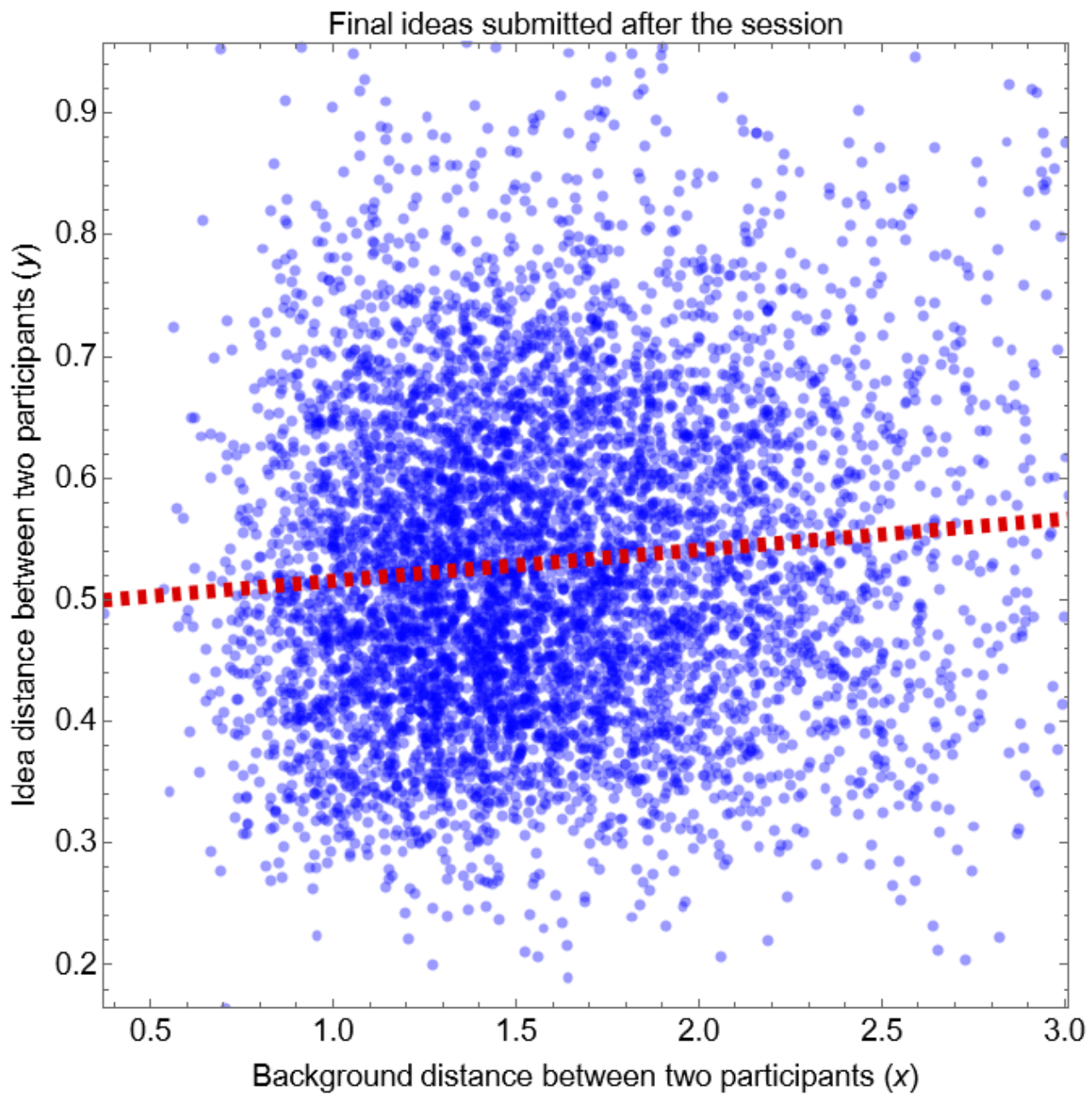

**Fig. S1.** Scatter plots showing positive correlations between the background distance between two participants ($x$) and the idea distance between them ($y$). Data were collected from the sessions of the "FC-SL" experiment. Dashed red lines show the results of linear model fitting. (A) Data collected during the first 10-day experimental sessions ($y = 0.0608004\,x + 0.552831$; $p = 1.222\times10^{-24}$), in which the centroid of multiple daily idea vectors submitted by each participant on each day was used to characterize their overall daily idea. (B) Data collected from the final ideas submitted after the experimental sessions ($y = 0.025398\,x + 0.490235$; $p = 7.16088\times10^{-13}$), in which the centroid of multiple final idea vectors submitted by each participant was used to characterize their overall final idea. Both plots exhibit statistically significant positive correlations, indicating the positive effects of the functional diversity of participants on idea diversification in their collective ideation processes.

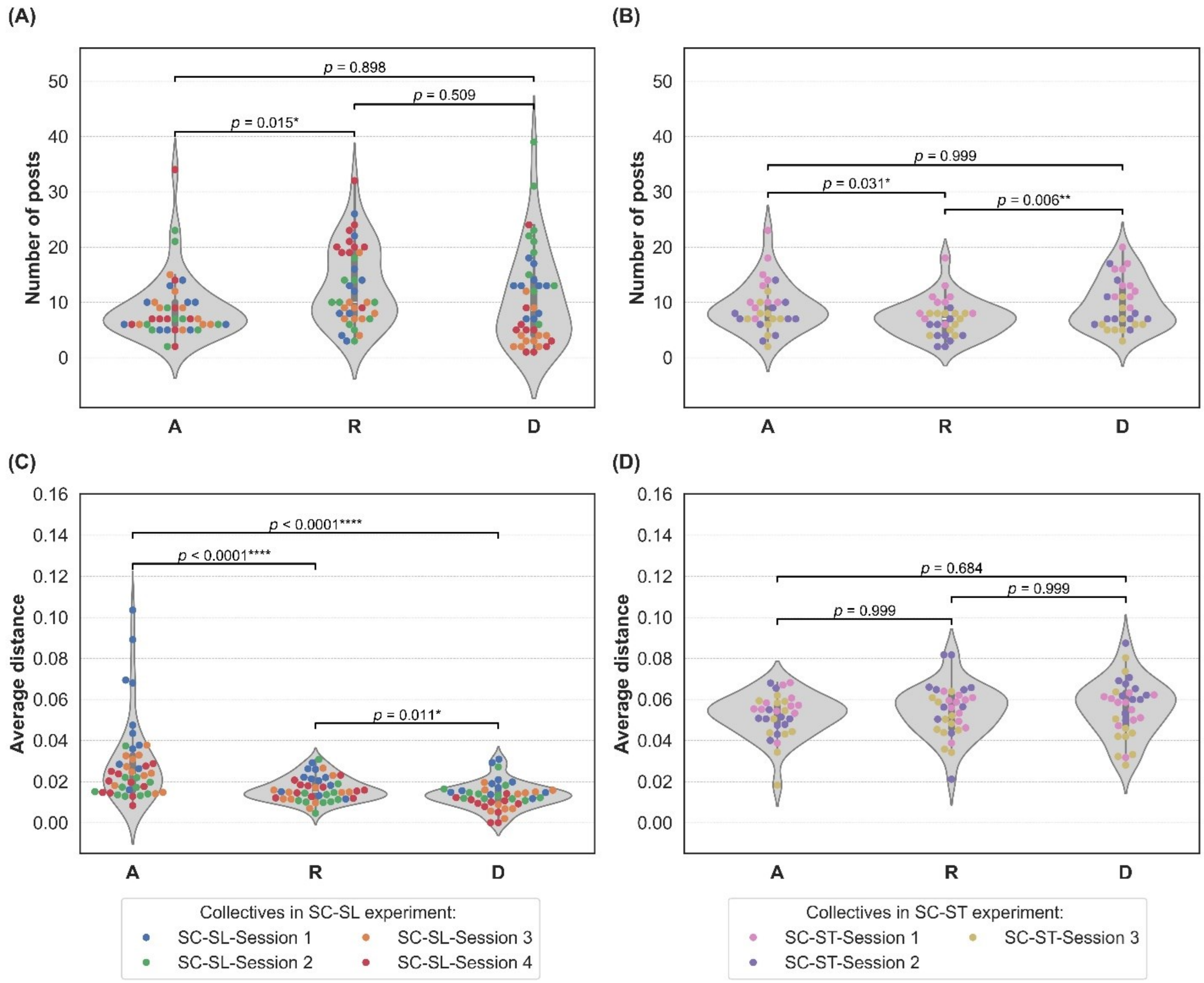

**Fig. S2.** Distribution of numbers of daily posts and average distances between ideas from Day 1 to Day 10*, including the outlier data points in (C)* (otherwise the same as Fig. 2 in the main text). (**A**) Comparison of numbers of daily posts among three background distribution conditions in the SC-SL experiment. (**B**) Comparison of numbers of daily posts among three background distribution conditions in the SC-ST experiment. (**C**) Comparison of average distances between ideas among three background distribution conditions in the SC-SL experiment. (**D**) Comparison of average distances between ideas among three background distribution conditions in the SC-ST experiment. Each dot in these violin plots represents the number of posts (for (**A**) and (**B**)) and the average value of distances between ideas (for (**C**) and (**D**)) of a collective for a single working day. The $p$-value annotation legend is as follows. *: $0.01 < p <= 0.05$, **: $0.001 < p <= 0.01$, ****: $p <= 0.0001$. The Wilcoxon signed-rank test with Bonferroni correction was used for all tests.

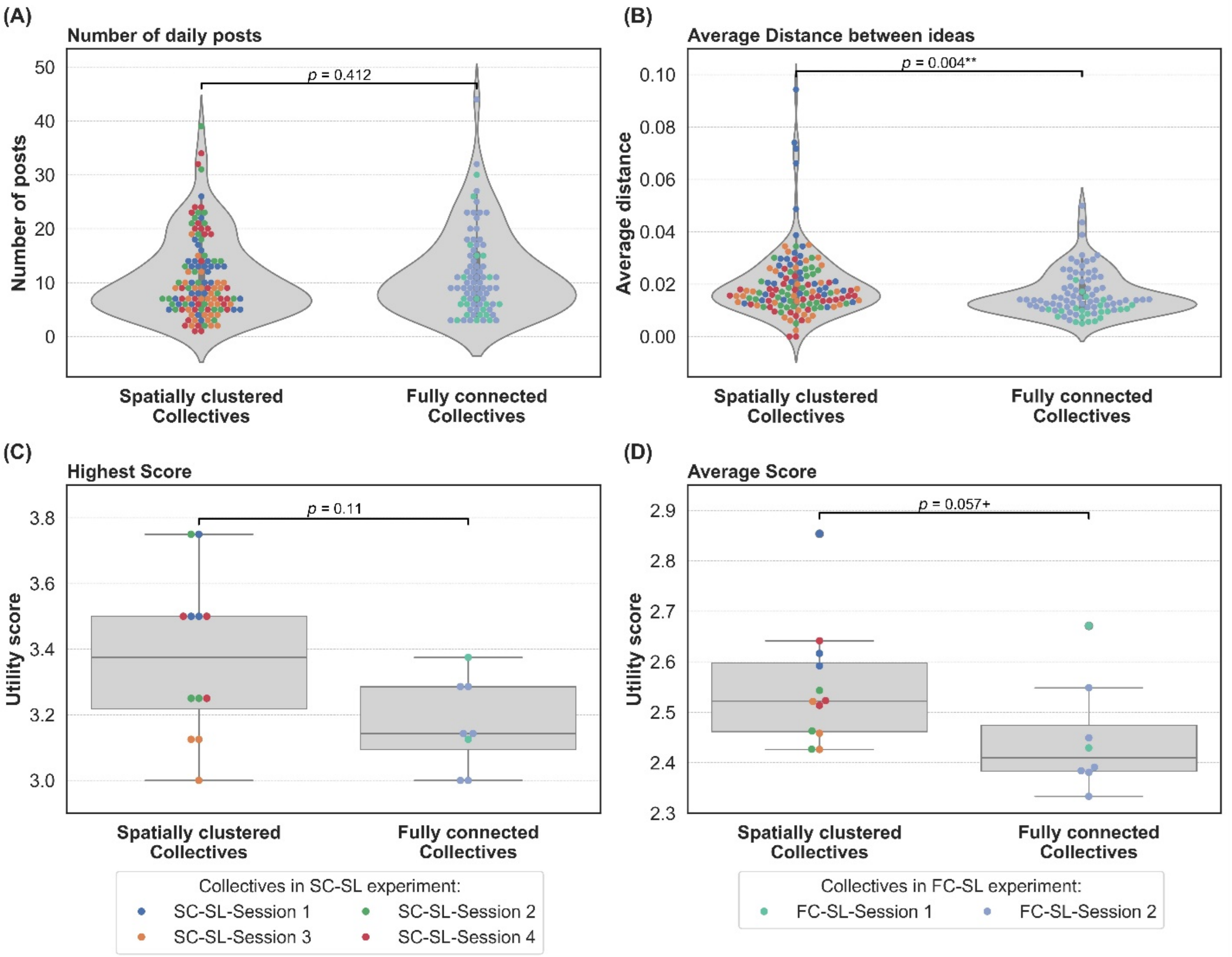

**Fig. S3.** Comparison of results between spatially clustered collectives in the SC-SL experiment and fully connected collectives in the FC-SL experiment*, including the outlier data points in (B)* (otherwise the same as Fig. 4 in the main text). (**A**) Number of daily posts. (**B**) Average distances between ideas. Each dot in these violin plots represents the number of posts (for (**A**)) and the average value of distances between ideas (for (**B**)) of a collective for a single working day. (**C**) Highest score of final ideas. (**D**) Average score of final ideas. The *p*-value annotation legend is as follows. +: $0.05 < p <= 0.1$, **: $0.001 < p <= 0.01$. The two-sided Mann-Whitney U test was used for all tests.

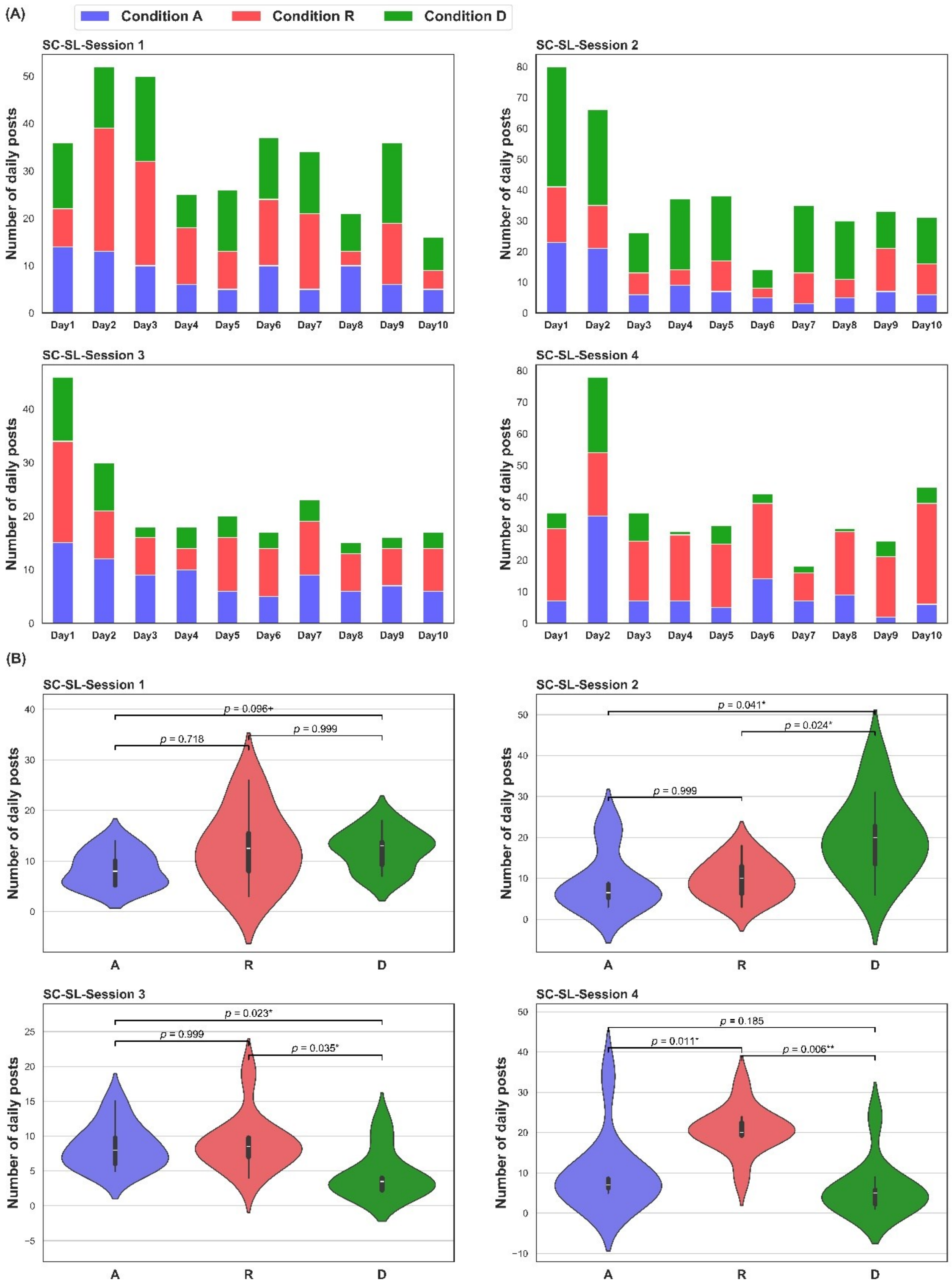

**Fig. S4.** Numbers of daily posts for each session in the SC-SL experiment. (**A**) Time series of numbers of daily posts from Day 1 to Day 10. (**B**) Comparison of distributions of numbers of daily posts among three collectives. The *p*-value annotation legend is as follows. +: $0.05 < p <= 0.1$, *: $0.01 < p <= 0.05$, **: $0.001 < p <= 0.01$, ****: $p <= 0.0001$. The two-sided Mann-Whitney U test with Bonferroni correction was used for all tests.

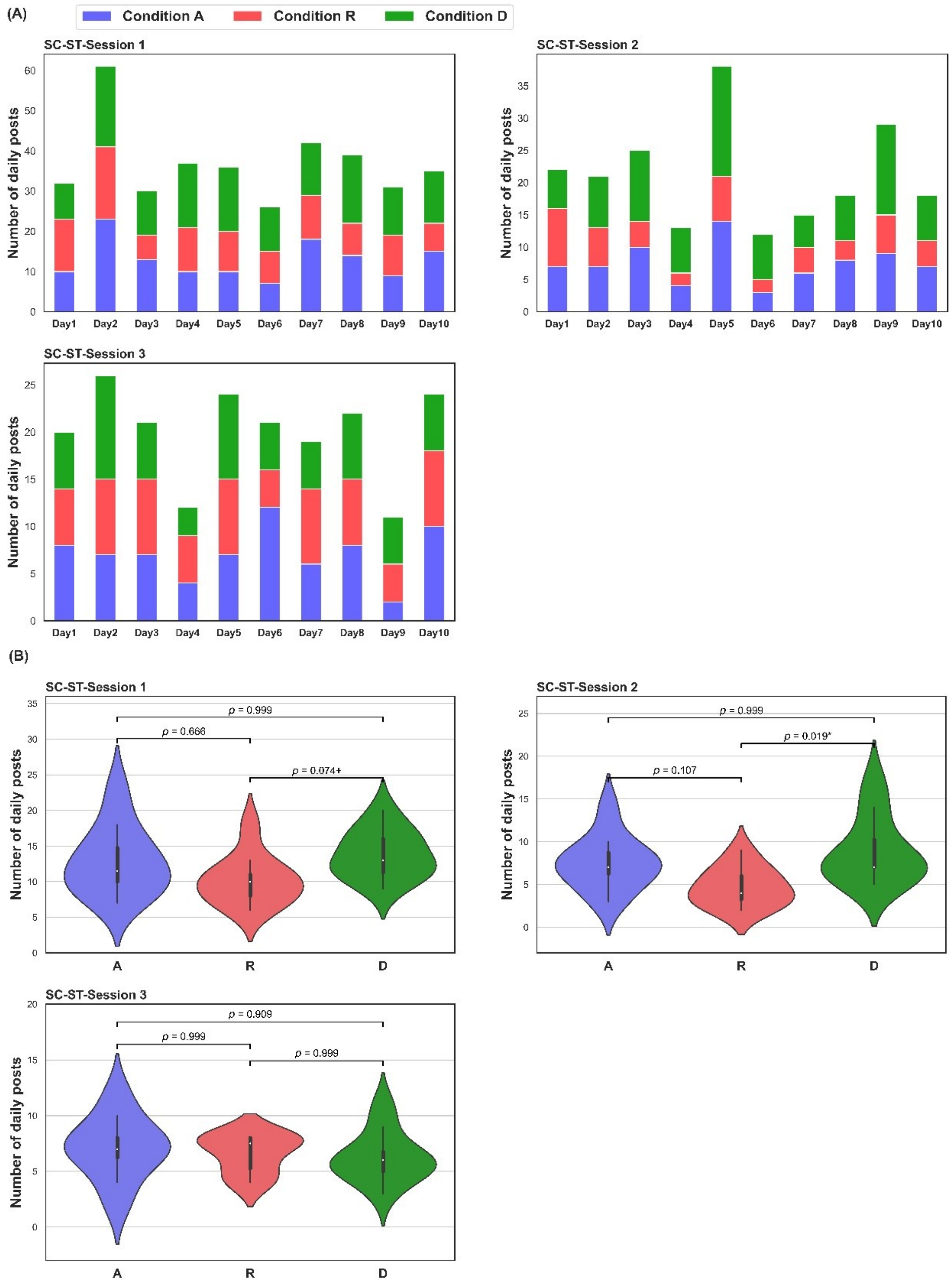

**Fig. S5.** Numbers of daily posts for each session in the SC-ST experiment. (**A**) Time series of numbers of daily posts from Day 1 to Day 10. (**B**) Comparison of distributions of numbers of daily posts among three collectives. The $p$-value annotation legend is as follows. +: $0.05 < p <= 0.1$, *: $0.01 < p <= 0.05$. The two-sided Mann-Whitney U test with Bonferroni correction was used for all tests.

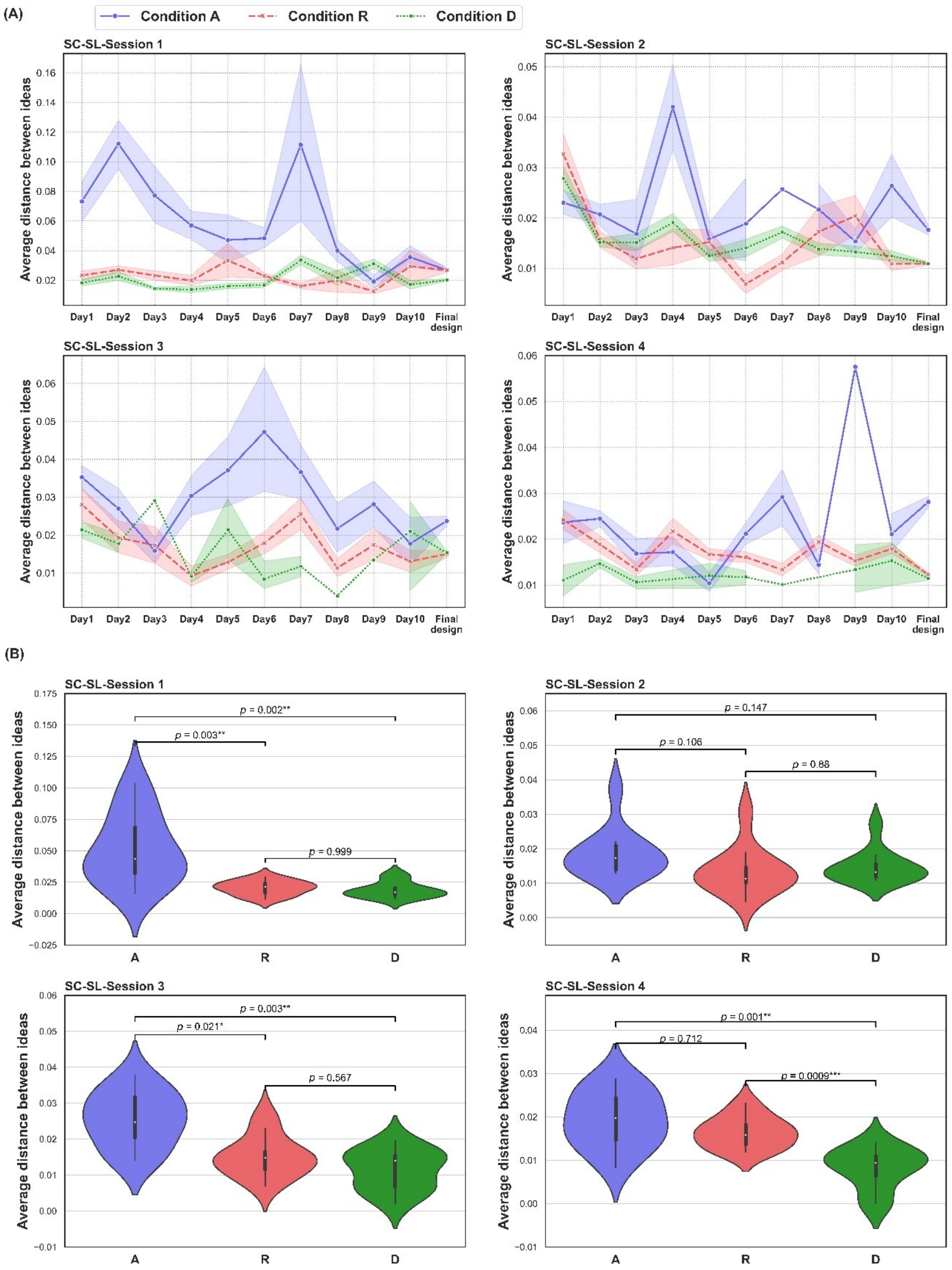

**Fig. S6.** Average distance between ideas for each session in the SC-SL experiments. (**A**) Time series of average distance between ideas from Day 1 to Day 10. (**B**) Comparison of distributions of average distance between ideas among three collectives. The *p*-value annotation legend is as follows. *: $0.01 < p <= 0.05$, **: $0.001 < p <= 0.01$, ***: $0.0001 < p <= 0.001$. The two-sided Mann-Whitney U test with Bonferroni correction was used for all tests.

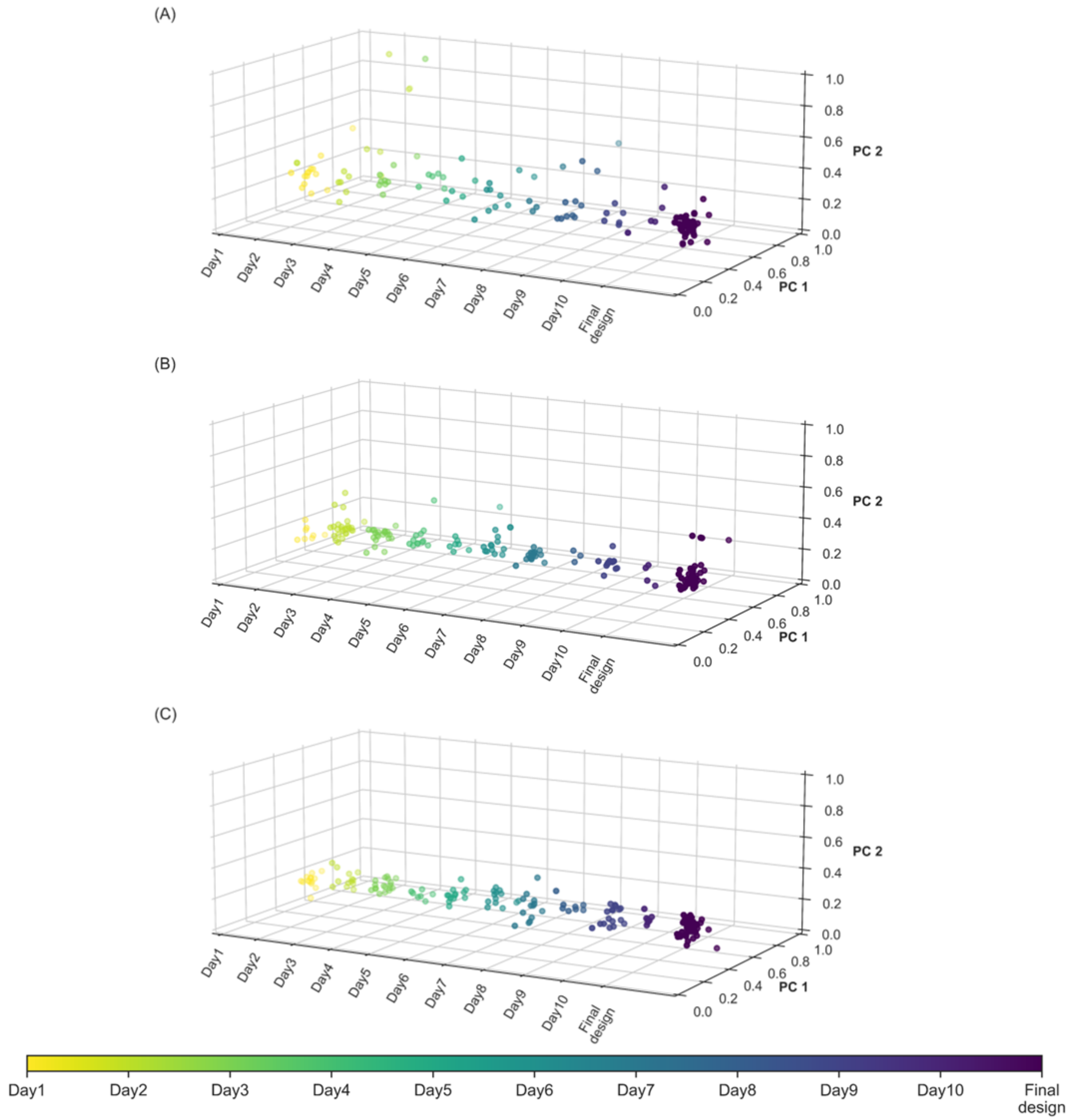

**Fig. S7.** Time series of Idea Cloud *(1)* of collectives in SC-SL Session 1. (**A**) Collective with Condition A. (**B**) Collective with Condition R. (**C**) Collective with Condition D.

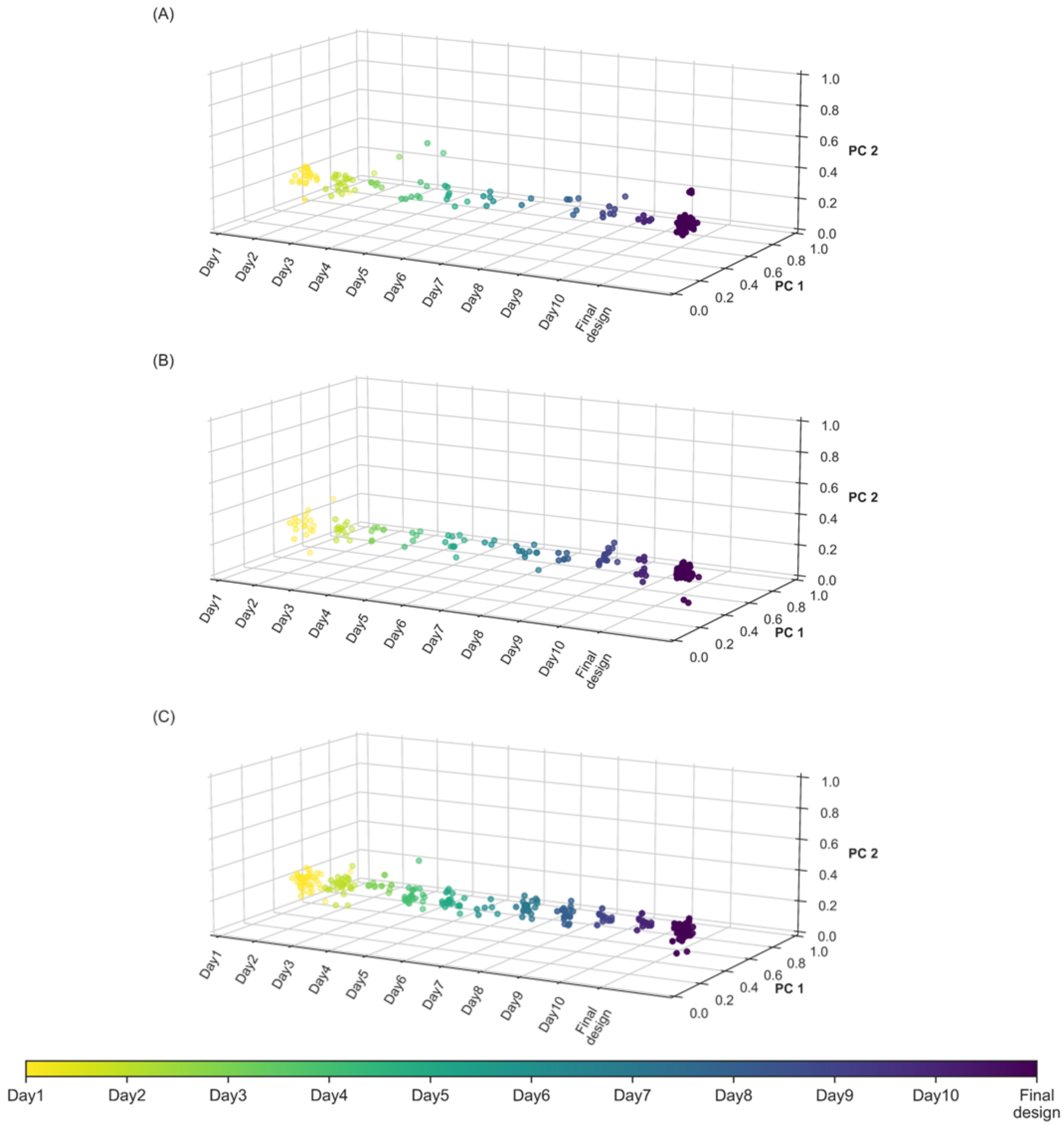

**Fig. S8.** Time series of Idea Cloud *(1)* of collectives in SC-SL Session 2. (**A**) Collective with Condition A. (**B**) Collective with Condition R. (**C**) Collective with Condition D.

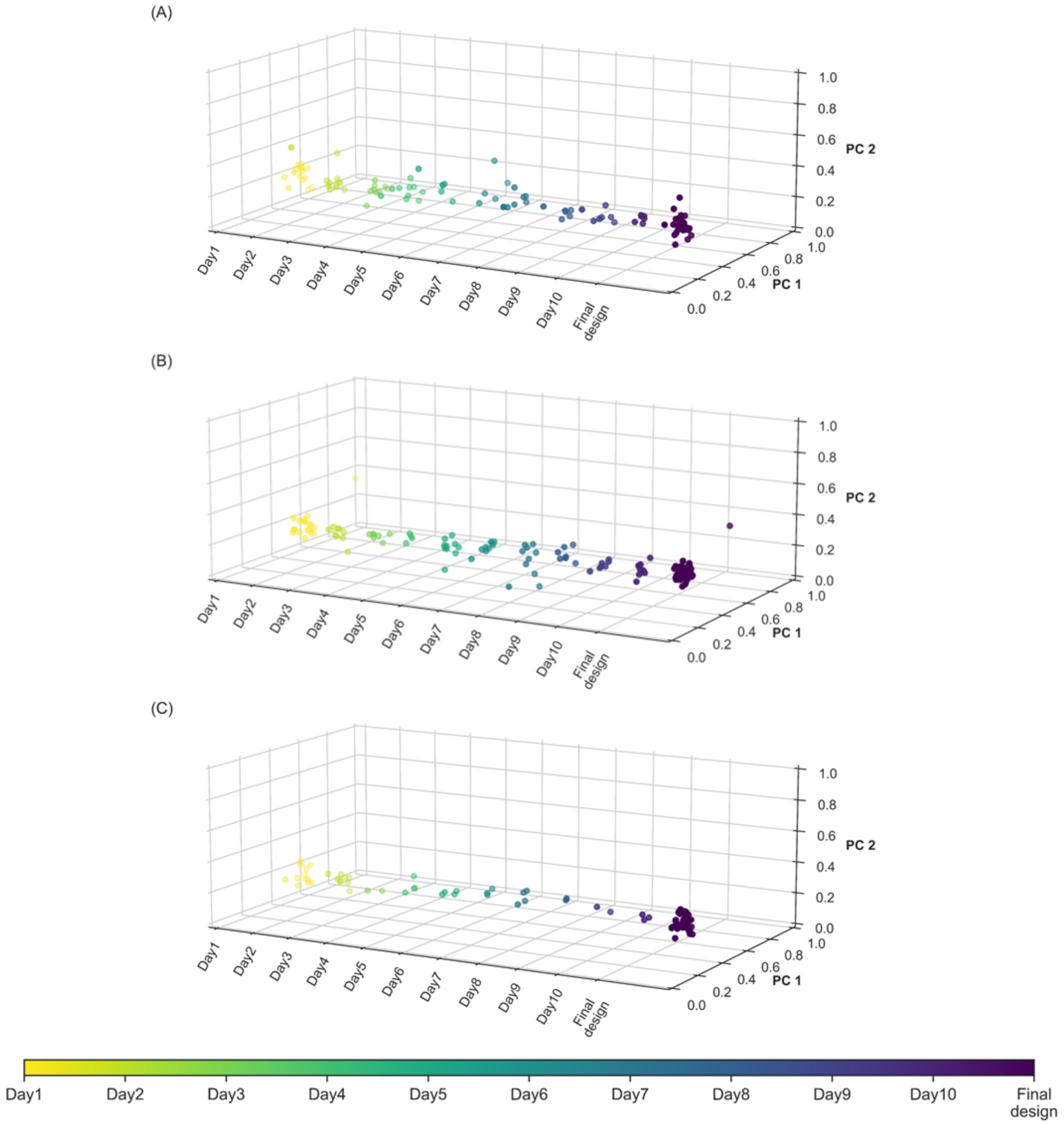

**Fig. S9.** Time series of Idea Cloud *(1)* of collectives in SC-SL Session 3. (**A**) Collective with Condition A. (**B**) Collective with Condition R. (**C**) Collective with Condition D.

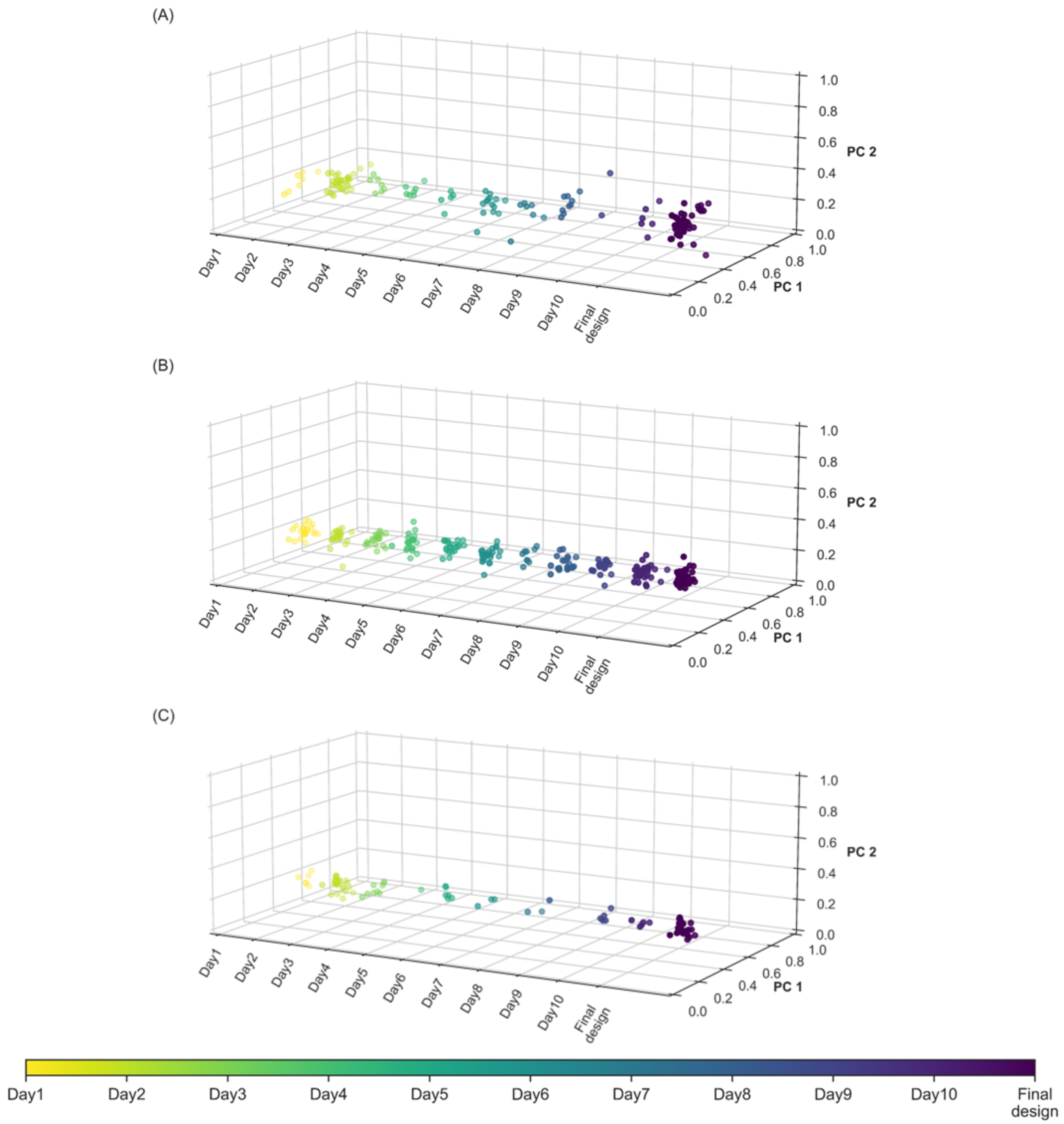

**Fig. S10.** Time series of Idea Cloud *(1)* of collectives in SC-SL Session 3. (**A**) Collective with Condition A. (**B**) Collective with Condition R. (**C**) Collective with Condition D.

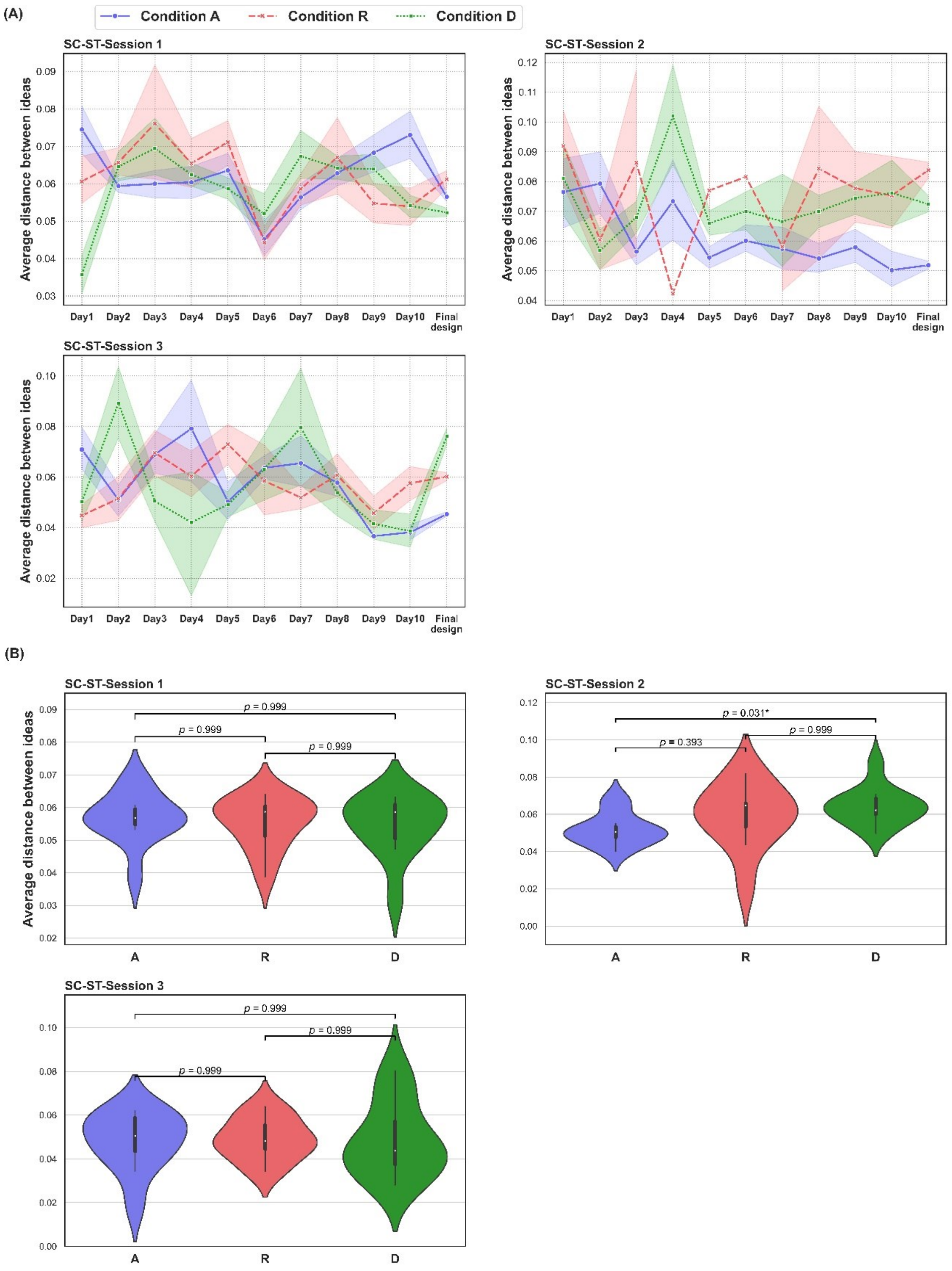

**Fig. S11.** Average distance between ideas for each session in the SC-SL experiment. (**A**) Time series of average distance between ideas from Day 1 to Day 10. (**B**) Comparison of average distance between ideas among three collectives. The *p*-value annotation legend is as follows. *: $0.01 < p <= 0.05$. The two-sided Mann-Whitney U test with Bonferroni correction was used for all tests.

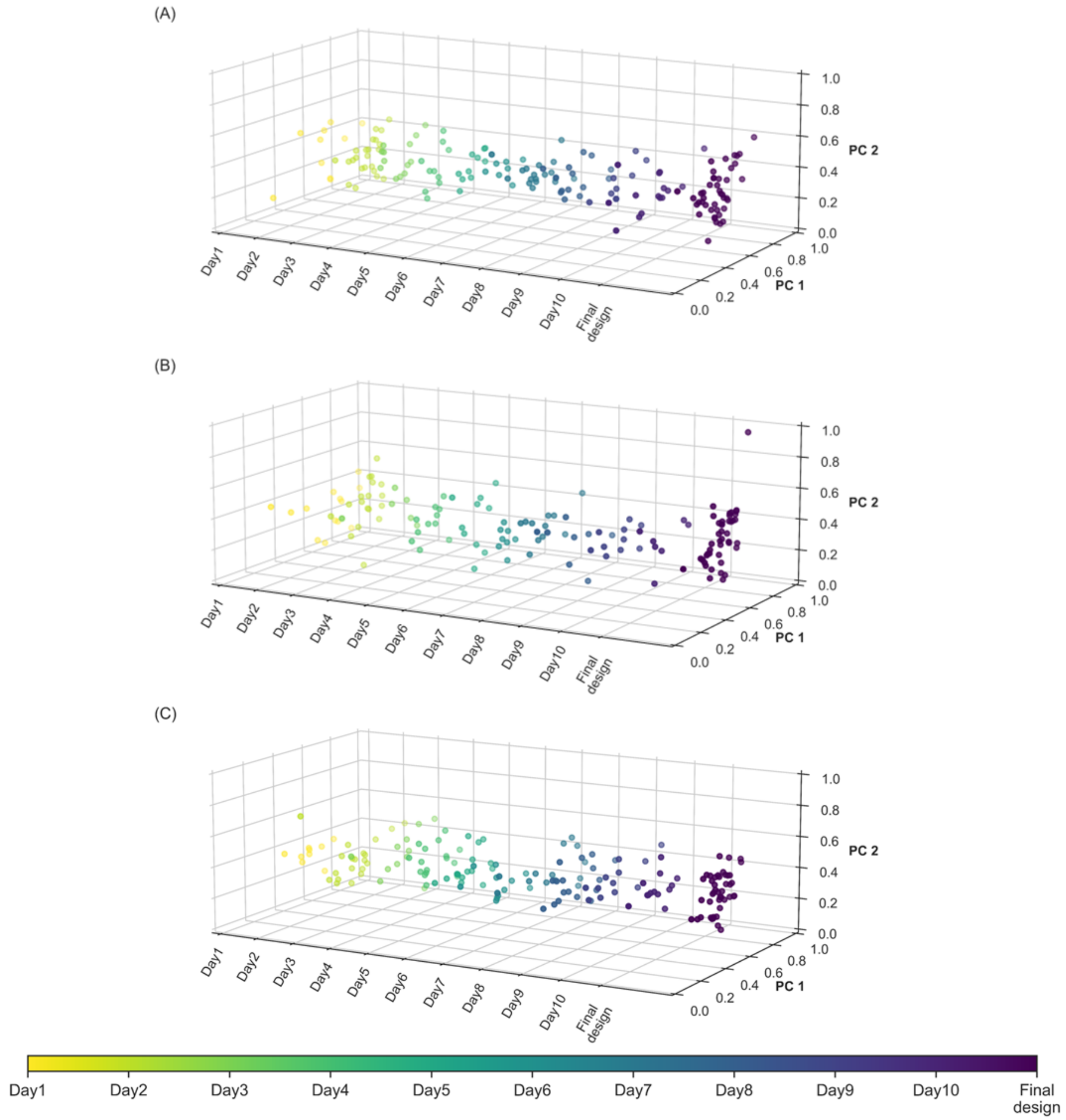

**Fig. S12.** Time series of Idea Cloud *(1)* of collectives in SC-ST Session 1. (**A**) Collective with Condition A. (**B**) Collective with Condition R. (**C**) Collective with Condition D.

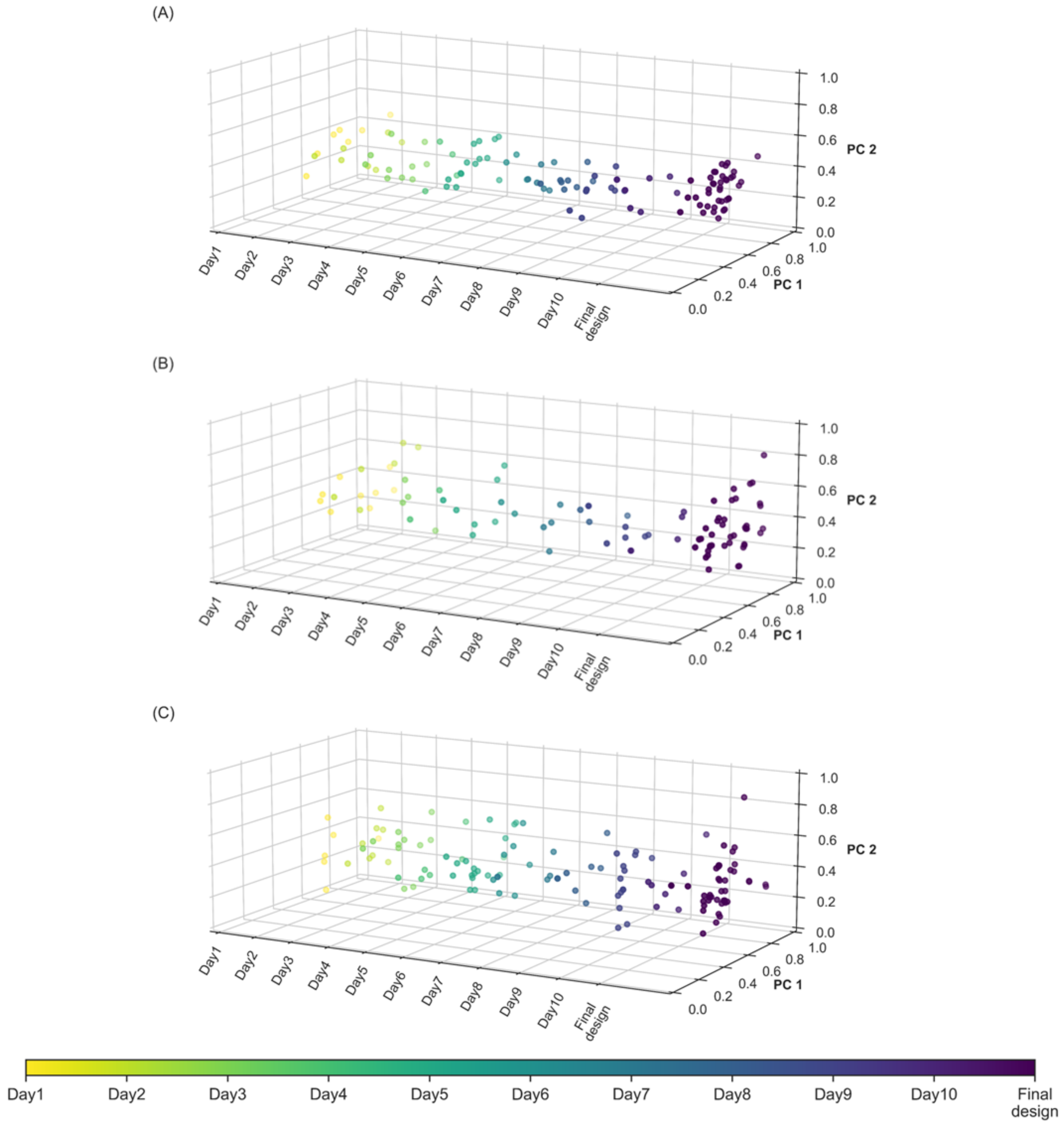

**Fig. S13.** Time series of Idea Cloud *(1)* of collectives in SC-ST Session 2. (**A**) Collective with Condition A. (**B**) Collective with Condition R. (**C**) Collective with Condition D.

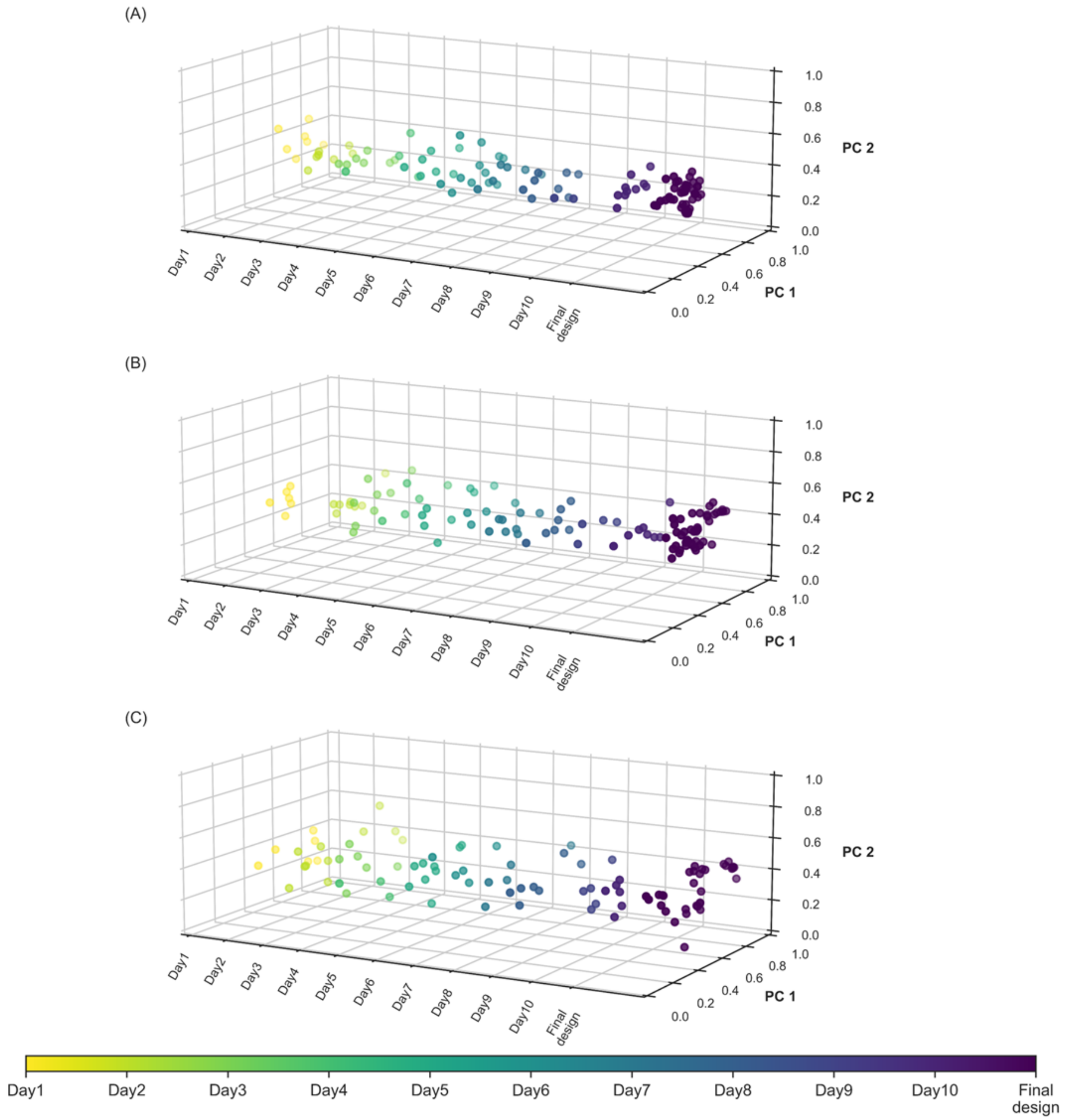

**Fig. S14.** Time series of Idea Cloud *(1)* of collectives in SC-ST Session 3. (**A**) Collective with Condition A. (**B**) Collective with Condition R. (**C**) Collective with Condition D.

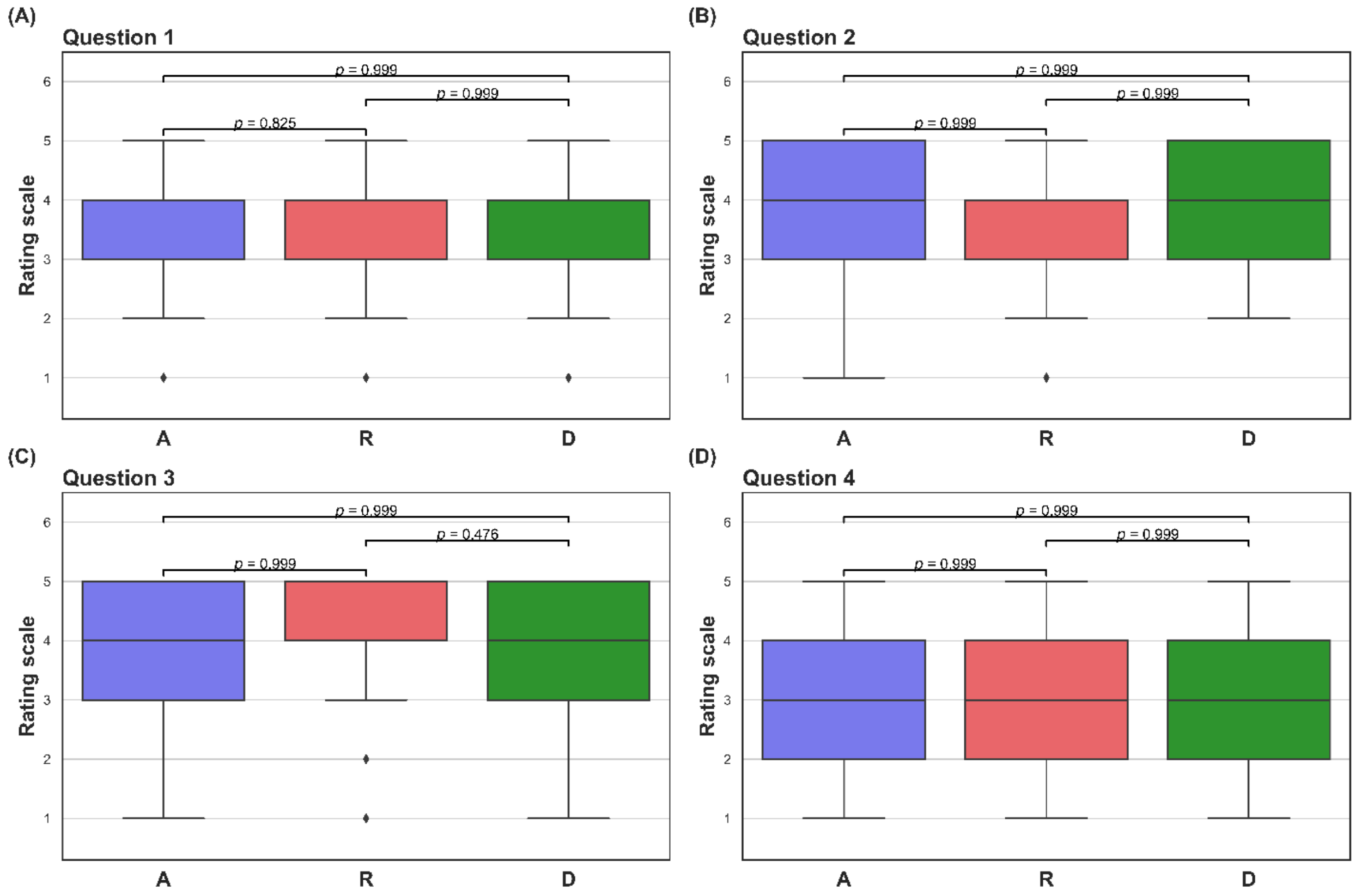

**Fig. S15.** Comparison of rating scale distributions in response to the end-of-the-session survey questions among three background distribution conditions for SC-SL sessions. (**A**) Question 1 about overall experience. (**B**) Question 2 about self-evaluated overall quality. (**C**) Question 3 about self-evaluated contribution. (**D**) Question 4 about learning experience. The two-sided Mann-Whitney U test with Bonferroni correction was used for all tests.

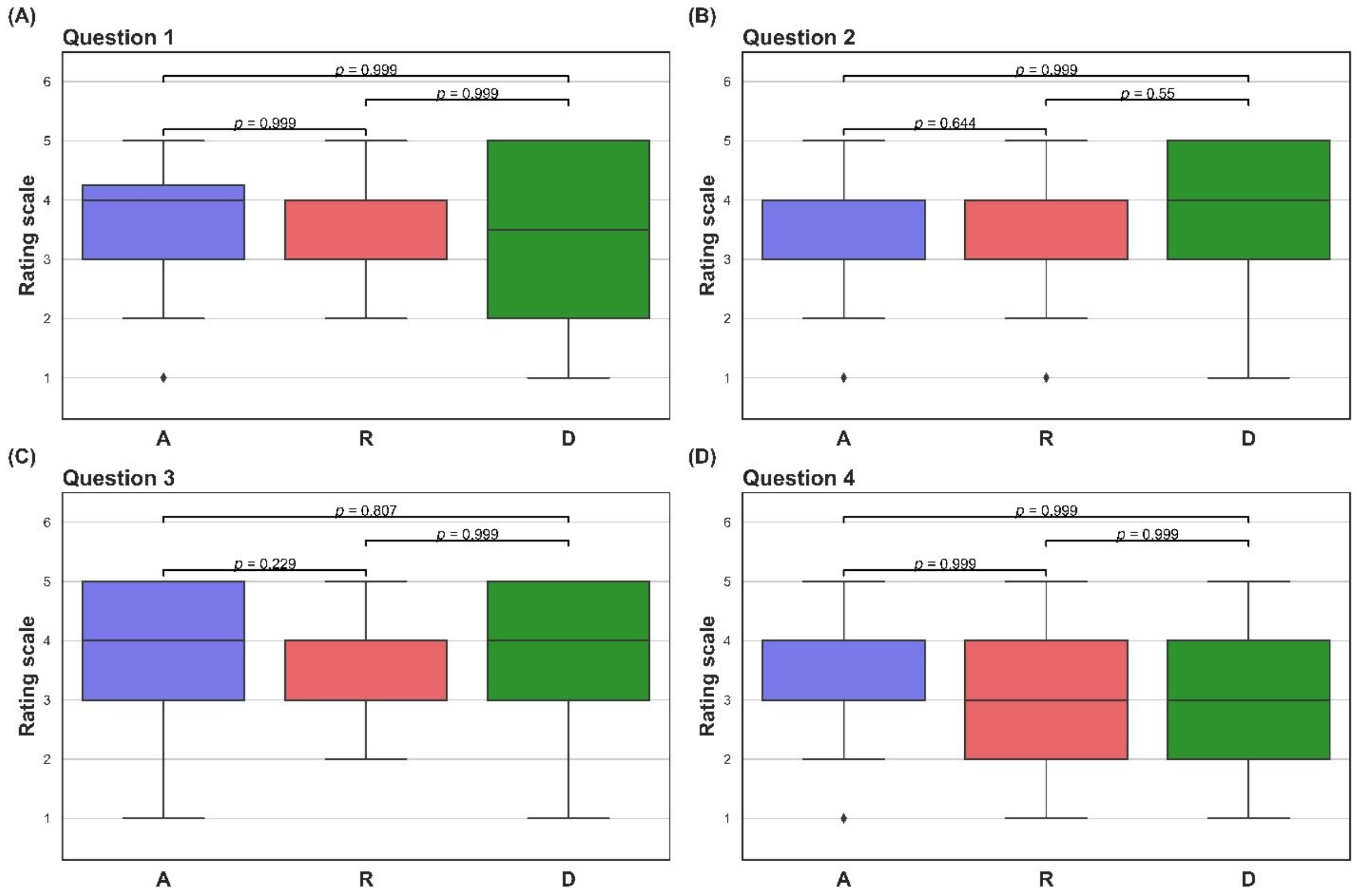

**Fig. S16.** Comparison of rating scale regarding end-of-the-session questions among three background distribution conditions for SC-ST sessions. (**A**) Question 1 about overall experience. (**B**) Question 2 about self-evaluated overall quality. (**C**) Question 3 about self-evaluated contribution. (**D**) Question 4 about learning experience. The two-sided Mann-Whitney U test with Bonferroni correction was used for all tests.